\documentclass[a4paper,11pt]{article}
\usepackage[dvipsnames]{xcolor}
\usepackage{pgfplots}
\usepackage{tikz-3dplot}
\usepackage{jheppub}
\usepackage[T1]{fontenc} 
\usepackage{amsmath}
\usepackage{amsthm}
\usepackage{amssymb}
\usepackage{physics}
\usepackage[compat=1.1.0]{tikz-feynman}
\usepackage[T1]{fontenc} 
\usepackage{tensor}
\usepackage{bm}
\usepackage{graphicx}
\usepackage[export]{adjustbox}
\usepackage{slashed}
\usepackage{stackrel,amssymb}
\usepackage{tikz}
\usepackage{mathtools}
\usetikzlibrary{arrows.meta}
\usepackage{multirow}
\usepackage{bm}
\usepackage{xcolor}
\usepackage{soul}
\usepackage{amsfonts}
\usepackage{comment}
\tikzset{>={Latex[scale=1.1]}}
\usetikzlibrary{arrows.meta,
	bending,
	decorations.markings, decorations.text}
\newcommand{\overbar}[1]{\mkern 1.5mu\overline{\mkern-1.5mu#1\mkern-1.5mu}\mkern 1.5mu}
\newcommand{\ee}{\mathrm{e}}
\newcommand{\ii}{\mathrm{i}}

\makeatletter
\newcommand*{\letterdef@}{}
\newcommand*{\letterdef}[3]{%
	\def\letterdef@##1{\expandafter\newcommand\csname #1\endcsname{#2{##1}}}%
	\@tfor\@tempa :=#3\do{\expandafter\letterdef@\expandafter{\@tempa}}}
\makeatother
\letterdef{c#1}{\mathcal}{ABCDEFGHIJKLMNOPQRSTUVWXYZ}
\letterdef{rm#1}{\mathrm} {dDeimM} 

\makeatletter
\let\save@mathaccent\mathaccent
\newcommand*\if@single[3]{%
	\setbox0\hbox{${\mathaccent"0362{#1}}^H$}%
	\setbox2\hbox{${\mathaccent"0362{\kern0pt#1}}^H$}%
	\ifdim\ht0=\ht2 #3\else #2\fi
}
\newcommand*\rel@kern[1]{\kern#1\dimexpr\macc@kerna}
\newcommand*\widebar[1]{\@ifnextchar^{{\wide@bar{#1}{0}}}{\wide@bar{#1}{1}}}
\newcommand*\wide@bar[2]{\if@single{#1}{\wide@bar@{#1}{#2}{1}}{\wide@bar@{#1}{#2}{2}}}
\newcommand*\wide@bar@[3]{%
	\begingroup
	\def\mathaccent##1##2{%
		\let\mathaccent\save@mathaccent
		\if#32 \let\macc@nucleus\first@char \fi
		\setbox\z@\hbox{$\macc@style{\macc@nucleus}_{}$}%
		\setbox\tw@\hbox{$\macc@style{\macc@nucleus}{}_{}$}%
		\dimen@\wd\tw@
		\advance\dimen@-\wd\z@
		\divide\dimen@ 3
		\@tempdima\wd\tw@
		\advance\@tempdima-\scriptspace
		\divide\@tempdima 10
		\advance\dimen@-\@tempdima
		\ifdim\dimen@>\z@ \dimen@0pt\fi
		\rel@kern{0.6}\kern-\dimen@
		\if#31
		\overline{\rel@kern{-0.6}\kern\dimen@\macc@nucleus\rel@kern{0.4}\kern\dimen@}%
		\advance\dimen@0.4\dimexpr\macc@kerna
		\let\final@kern#2%
		\ifdim\dimen@<\z@ \let\final@kern1\fi
		\if\final@kern1 \kern-\dimen@\fi
		\else
		\overline{\rel@kern{-0.6}\kern\dimen@#1}%
		\fi
	}%
	\macc@depth\@ne
	\let\math@bgroup\@empty \let\math@egroup\macc@set@skewchar
	\mathsurround\z@ \frozen@everymath{\mathgroup\macc@group\relax}%
	\macc@set@skewchar\relax
	\let\mathaccentV\macc@nested@a
	\if#31
	\macc@nested@a\relax111{#1}%
	\else
	\def\gobble@till@marker##1\endmarker{}%
	\futurelet\first@char\gobble@till@marker#1\endmarker
	\ifcat\noexpand\first@char A\else
	\def\first@char{}%
	\fi
	\macc@nested@a\relax111{\first@char}%
	\fi
	\endgroup
}
\makeatother
\newdimen\tableauside\tableauside=1.0ex
\newdimen\tableaurule\tableaurule=0.4pt
\newdimen\tableaustep
\def\phantomhrule#1{\hbox{\vbox to0pt{\hrule height\tableaurule
			width#1\vss}}}
\def\phantomvrule#1{\vbox{\hbox to0pt{\vrule width\tableaurule
			height#1\hss}}}
\def\sqr{\vbox{%
		\phantomhrule\tableaustep
		\hbox{\phantomvrule\tableaustep\kern\tableaustep\phantomvrule\tableaustep}%
		\hbox{\vbox{\phantomhrule\tableauside}\kern-\tableaurule}}}
\def\squares#1{\hbox{\count0=#1\noindent\loop\sqr
		\advance\count0 by-1 \ifnum\count0>0\repeat}}
\def\tableau#1{\vcenter{\offinterlineskip
		\tableaustep=\tableauside\advance\tableaustep by-\tableaurule
		\kern\normallineskip\hbox
		{\kern\normallineskip\vbox
			{\gettableau#1 0 }%
			\kern\normallineskip\kern\tableaurule}%
		\kern\normallineskip\kern\tableaurule}}
\def\gettableau#1 {\ifnum#1=0\let\next=\null\else
	\squares{#1}\let\next=\gettableau\fi\next}
\newcommand{\Yfund}{\tableau{1}}
\newcommand{\Ysymm}{\tableau{2}}
\newcommand{\Yasymm}{\tableau{1 1}}

\tdplotsetmaincoords{60}{115}
\pgfplotsset{compat=newest}

\author[a,b]{M. Bill\`o,}
\author[c]{L. Griguolo,}
\author[d,b]{A. Lerda,}
\author[c,e]{and A. Testa\,}

\affiliation[a]{Universit\`a di Torino, Dipartimento di Fisica,\\ Via P. Giuria 1, I-10125 Torino, Italy}
	\affiliation[b]{INFN, Sezione di Torino,	\\Via P. Giuria 1, I-10125 Torino, Italy}
	\affiliation[c]{Dipartimento SMFI, Universit\`a di Parma and INFN, Gruppo Collegato di Parma,
	\\ Viale G.P. Usberti 7/A, I-43100 Parma, Italy}
\affiliation[d]{Universit\`a del Piemonte Orientale,
			Dipartimento di Scienze e Innovazione Tecnologica\\
			Viale T. Michel 11, I-15121 Alessandria, Italy}
	\affiliation[e]{Institut de Physique Th\'eorique\footnote{Unit\'e Mixte de Recherche 3681 du CNRS}, Universit\'e Paris Saclay, CNRS,  91191 Gif-sur-Yvette, France}
\emailAdd{marco.billo@unito.it}
\emailAdd{luca.griguolo@unipr.it}
\emailAdd{lerda@to.infn.it}
\emailAdd{alessandro.testa@unipr.it
}

\title{Correlators in non-conformal $\mathcal{N}=2$ gauge theories from localization}

\abstract{We study two-point correlation functions of chiral/anti-chiral operators in SU($N$) $\mathcal{N}=2$ gauge theories with massless hyper-multiplets in a representation  $\cR$ associated with a non-vanishing $\beta$-function. Using supersymmetric localization on the four-sphere $S^4$, these observables can be evaluated by matrix correlators of normal ordered operators. We show that, within a specific regime of validity,   standard perturbative calculations based on Feynman diagrams and renormalization procedures in flat space perfectly match the localization predictions up to two loops generalizing and extending previous results. Our analysis highlights that non-trivial interference effects  between evanescent terms and the UV poles of the bare coupling are essential for the agreement. On the matrix model side, we employ a direct diagrammatic procedure akin to the Feynman diagram expansion on the field theory description; this simplifies the comparison for generic operators and general matter representation $\cR$.}


\begin{document}
\maketitle
\flushbottom

\section{Introduction}
Supersymmetric localization \cite{Pestun:2016zxk} is a powerful technique for studying quantum field theories in perturbative and non-perturbative regimes. Under suitable assumptions, it allows the evaluation of various protected observables defined on compact space-times by means of matrix-model averages \cite{Pestun:2007rz}.

In four-dimensional $\mathcal{N}=4$ and \emph{superconformal} $\mathcal{N}=2$ gauge theories, supersymmetric localization on the four-sphere $S^4$ has been extensively used to study families of BPS local operators \cite{Baggio:2014ioa,Baggio:2014sna,Gerchkovitz:2016gxx,Rodriguez-Gomez:2016cem,Rodriguez-Gomez:2016ijh,Billo:2017glv,Billo:2018oog,Galvagno:2020cgq,Beccaria:2020hgy,Fiol:2021icm}, extended probes such as supersymmetric Wilson loops \cite{Pestun:2007rz,Drukker:2007qr,Bassetto:2008yf,Pestun:2009nn,Giombi:2009ds,Andree:2010na,Billo:2019fbi,Beccaria:2021vuc} and interfaces \cite{Komatsu:2020sup,Beccaria:2022bjo}, Bremsstrahlung functions \cite{Correa:2012at,Bonini:2015fng,Fiol:2015spa,Mitev:2015oty,Bianchi:2019dlw} and the so-called ``integrated correlators'' \cite{Binder:2019jwn,Chester:2019jas,Chester:2020dja,Dorigoni:2021guq,Paul:2022piq,Alday:2023pet,Brown:2024tru,Behan:2023fqq,Billo:2023kak,Pini:2024uia,Billo:2024ftq,Pini:2024zwi,DeLillo:2025hal}. Exploiting conformal symmetry one can naturally extend the results on $S^4$ to flat space, thereby allowing for the use of the matrix integrals generated by localization to probe the strong coupling dynamics of superconformal gauge theories in $\mathbb{R}^4$ \cite{Beccaria:2020hgy,Beccaria:2021ism,Beccaria:2021hvt,Billo:2021rdb,Billo:2022gmq,Billo:2022fnb,Beccaria:2022ypy,Beccaria:2023kbl,Billo:2022xas,Pini:2023svd,Pini:2023lyo,Korchemsky:2025eyc} and, in certain cases, also their dual holographic descriptions.

However, when the theory contains dimensionful parameters, such as a mass term in $\mathcal{N}=2^*$ Super Yang-Mills (SYM), or a scale dynamically generated by dimensional transmutation, conformal symmetry is broken. This breaking leads to a non-trivial separation between short- and long-distance physics, and consequently the connection of observables computed on $S^4$ and their flat-space counterparts becomes less clear. For instance, in massive theories the observables on the sphere depend explicitly on the dimensionless combination $RM$, where $R$ is the radius of $S^4$ and $M$ is the mass scale appearing in the Lagrangian; this dependence typically differs from that of the corresponding flat-space observables already at perturbative level, as explicitly shown in \cite{Belitsky:2020hzs} for the $\frac{1}{2}$-BPS circular Wilson loop. In asymptotically free theories, where conformal symmetry is dynamically broken via a non-zero $\beta$-function, the situation is more subtle and, although the presence of a running coupling introduces a scale dependence, certain localization results on $S^4$ still retain a meaningful connection to flat-space physics.

A first preliminary attempt to investigate whether the correlation functions of gauge-invariant local operators in $\mathbb{R}^4$ could be obtained using localization techniques in $S^4$ was made in \cite{Billo:2019job} at the two-loop order in non-conformal $\mathcal{N}=2$ SQCD. There it was found that the results in $\mathbb{R}^4$ matched those of the matrix model in $S^4$ only for a class of suitably normalized correlators that are insensitive to the non-zero $\beta$-function of the theory. A more complete and exhaustive analysis has been recently performed in \cite{Billo:2023igr,Billo:2024fst,Billo:2024hvf} for the expectation value of $\frac{1}{2}$-BPS Wilson loops in non-conformal $\mathcal{N}=2$ gauge theories with massless hyper-multiplets in a generic representation $\cR$. In these papers it has been shown that, within a specific regime of validity, there is a perfect consistency between the diagrammatic results in flat space and the localization predictions up to three loops. The reason is that at a given order in perturbation theory there are \emph{evanescent} contributions that vanish in four dimensions but which in non-conformal theories can interfere with the ultraviolet (UV) poles of the running coupling in dimensional regularization and give rise to finite terms at the next perturbative order. Taking these contributions into account a perfect match between the flat-space results and the sphere localization predictions is obtained even for theories with a non-vanishing $\beta$-functions, contrary to naive expectations. The goal of this work is to extend the analysis of \cite{Billo:2023igr,Billo:2024fst,Billo:2024hvf} and show that such agreement exists in full generality also for the correlators of local chiral and anti-chiral operators. More precisely, we prove that in $\mathcal{N}=2$ gauge theories with massless hyper-multiplets and  non-vanishing $\beta$-function, supersymmetric localization on $S^4$ allows us to compute two-point functions of local chiral/anti-chiral operators that are in complete agreement with the results on $\mathbb{R}^4$ in the perturbative regime
\begin{equation}
\Lambda \ll 1/R \ll \mu
\end{equation}
where $\Lambda$ is the dynamically generated energy scale at which the running coupling diverges, $R$ is the sphere radius and $\mu$ is the renormalization scale\,\footnote{Note that in this regime the running coupling is $\ll 1$ so that the non-perturbative instanton contributions are exponentially suppressed and can be neglected.}.
 
The paper is organized as follows. In Section\,\ref{sec:Section 2}, we present the perturbative calculations of the two-point correlators in flat space using Feynman diagrams. In particular, we generalize the analysis presented in \cite{Billo:2019job} to a generic representation $\cR$ of the gauge group and, importantly, show the emergence of new subtle interference effects due to evanescent terms which are essential to guarantee the agreement with the localization results. The matrix-model calculation is presented in Section\,\ref{sec:MM}, where we give the details for the derivation of the two-point correlators at the first non-trivial order in perturbation theory. We also show in full generality the perfect match with the flat-space results. Conclusions and future extensions of the present work are discussed in Section\,\ref{sec:conclusions}, where we comment on the possibility and interest of obtaining
the correlators at higher orders and at large $N$ using the matrix model. Finally, some calculational details are summarized in the appendices, where we collect various formulas and some explicit results for low dimensional operators.

\section{Field theory calculations}
\label{sec:Section 2}
We consider $\mathcal{N}=2$ SYM theories with gauge group SU($N$) and massless hypermultiplets in a generic representation $\cR$. In these theories,  classical conformal symmetry is dynamically broken at the quantum level by a one-loop exact $\beta$-function which, in $d = 4- 2 \epsilon$ dimensions, is
 \begin{equation}
	\label{beta}
			\beta(g_*) =-\epsilon g_* -\beta_0 \dfrac{g_*^3}{16\pi^2}~.
	\end{equation}		
Here $g_*$ is the renormalized coupling evaluated at the renormalization scale $1/\mu$ and
\begin{align}
\beta_0=2\big(N-i_{\cR}\big)~,
\label{beta0gen}
\end{align}
where $i_{\cR}$ is the Dynkin index of the representation $\cR$. Even if our analysis is completely general, we will sometimes consider specific $\mathcal{N}=2$ theories with $N_f$ fundamental, $N_s$ symmetric and $N_a$ anti-symmetric flavors. In this case 
\begin{equation}
\label{eq:Ris}
	\cR = N_f\, \Yfund \oplus N_s\, \Ysymm \oplus N_a\, \Yasymm~
\end{equation}
and
\begin{equation}
\label{beta0}
\beta_0=2N-N_f-N_a(N-2)-N_s(N+2)~.
\end{equation}		

Our goal is to study the correlators of the gauge-invariant  multi-trace operators
\begin{equation}
	\label{On}
O_{\vec{n}}(x) = \prod_{i=1}^{\ell}  g_B^{n_i}\tr\phi^{n_i}(x)
\end{equation}
where $\phi$ is the complex scalar of the vector multiplet, $\vec n = (n_1,n_2,\ldots,n_\ell)$ and $g_B$ is the {bare coupling}. These operators have a bare scaling dimension  
\begin{equation}
	\label{eq:conformal dimension}
	n = \sum_{i=1}^\ell n_i
\end{equation}
and a U(1) charge which, in our conventions, is equal to $n$.

Writing $\phi(x)=\phi^a(x) T^a$, where $T^a$ $(a = 1,\ldots, N ^2-1)$ are the generators of $\mathfrak{su}(N)$ in the fundamental representation\,\footnote{We normalize the generators such that $\tr T^a T^b = \frac 12 \,\delta_{ab}$.}, we see that (\ref{On}) becomes
\begin{equation}
	\begin{split}
		O_{\vec n}(x) =g_B^n\, R_{\vec n}^{\,a_1\ldots a_n}\, \phi^{a_1}(x)\ldots \phi^{a_n}(x)
	\end{split}
\end{equation} 
where the completely symmetric tensor $ R_{\vec n}^{\,a_1\ldots a_n}$ is 
\begin{equation}
	\label{Rtensor}
	R_{\vec{n}}^{\,a_1\dots a_n} 
= \tr T^{(a_1}\cdots T^{a_{n_1}}~
\tr T^{a_{n_1+1}}\cdots T^{a_{n_1+n_2}}\,\ldots\,
\tr T^{a_{n_1 + \ldots + n_{\ell-1}+1}}\cdots T^{a_n)}~.
\end{equation}
The anti-chiral operators $\widebar{O}_{\vec n}(x)$ are defined as in (\ref{On}) 
with $\phi(x)$ replaced by its complex conjugate $\widebar{\phi}(x)$.

\subsection{Two-point functions in perturbation theory}
The (dimensionally regularized) two-point functions of chiral/anti-chiral operators in theories with a non-vanishing $\beta$-function take the following form \cite{Billo:2019job}
\begin{equation}
	\label{OnOm}
	\begin{split}
		\big\langle\,O_{\vec n}(x) \,\widebar{O}_{\vec m}(0)\,\big\rangle = G_{\vec{n},\vec{m}}(g_B,x,\epsilon)\,\Delta^n(x,\epsilon)\, \delta_{nm}  \ 
	\end{split}
\end{equation} 
where $\Delta(x,\epsilon)$ is the massless scalar propagator in $d=4-2\epsilon$ dimensions
\begin{equation}
	\label{massless prop}
	\Delta(x,\epsilon) = \int\dfrac{\dd^d p}{\left(2\pi\right)^d} \dfrac{e^{\ii p\cdot x}}{p^2} =  \dfrac{\Gamma(1-\epsilon)}{4\pi\left(\pi x^2\right)^{1-\epsilon}} \ .
\end{equation}
In perturbation theory, the function $G_{\vec{n},\vec{m}}$ can be expanded in powers of $g_B$ according to
 \begin{equation}
	\label{calG}
	G_{\vec{n},\vec{m}}(g_B,x,\epsilon)= g_B^{2n}\,G^{(0)}_{\vec{n},\vec{m}}\Big[1+\sum_{k=1}^{\infty}g_B^{2k} \,
	\cG^{(k)}_{\vec{n},\vec{m}}(x,\epsilon)\Big]
\end{equation} 
where $G^{(0)}_{\vec{n},\vec{m}}$ is the tree-level term and $\cG^{(k)}_{\vec{n},\vec{m}}$ 
is the contribution at $k$ loops.
When the notation is not ambiguous, we simply write $\cG^{(k)}$ rather than $\cG^{(k)}_{\vec{n},\vec{m}}$ and always understand the factor $\delta_{nm}$ that enforces the U(1) charge conservation. 

The perturbative expansion (\ref{calG}) can be efficiently organized by decomposing the loop corrections $\cG^{(k)}$ as
\begin{equation}
	\label{W2kis}
	\cG^{(k)} = \cG^{(k)}_{\cR } +\cG^{(k)}_{\mathrm{v.m.}} 
\end{equation}
where $\cG^{(k)}_{\cR }$ accounts for diagrams with internal hyper-multiplet lines in the representation $\cR$, while $\cG^{(k)}_{\mathrm{v.m.}}$ captures the contributions coming from diagrams with internal vector-multiplet lines only. Of course this decomposition also holds for $\mathcal{N}=4$ SYM, which can be seen as a $\mathcal{N}=2$ theory with an adjoint hypermultiplet. So, in this case we can write (\ref{W2kis}) with $\cR=\mathrm{adj}$. On the other hand, it is well-known that in $\cN=4$ SYM the loop corrections identically vanish
in $d=4$. This means that\,\footnote{See Ref.\,\cite{Erickson:2000af} for an example of the relation (\ref{calg N=4}) in the calculation of supersymmetric Wilson loops. See also Ref.\,\cite{Bianchi:2023llc} where the two-point correlator of local operators of dimension 2 in $\mathcal{N}=4$ SYM theory has been computed in momentum space in $d=4-2\epsilon$ dimensions. We have checked that the $\epsilon$-term of our one-loop calculation given in (\ref{eq:internal exchange def}) is fully consistent with what reported in \cite{Bianchi:2023llc}.}
\begin{equation}
	\label{calg N=4}
 \cG^{(k)}_{\mathrm{v.m.}} =  - \cG^{(k)}_{\mathrm{adj}}   + \delta^\prime	\cG^{(k)}
\end{equation}
where $\cG^{(k)}_{\mathrm{adj}}$ describes the contributions of diagrams with internal hyper-multiplet lines in the adjoint representation, while $\delta^\prime\cG^{(k)}$ vanishes for $d\to 4$ and thus can be expanded in positive powers of $\epsilon$: it is an \emph{evanescent} term.  

In a generic $\cN=2$ gauge theory, the vector-multiplet contribution $\cG^{(k)}_{\mathrm{v.m.}} $ is in common with $\cN=4$ SYM, and so we can use (\ref{calg N=4}) in (\ref{W2kis}) to get
\begin{equation}
	\label{calG def}
	\cG^{(k)} =  \cG^{(k)}_{\cR }  -  \cG^{(k)}_{\mathrm{adj}} + \delta^\prime	\cG^{(k)} \,\equiv\, \cG^{\prime\, (k)}  + \delta^\prime	\cG^{(k)} ~.
\end{equation}
Note that $\cG^{\prime\,(k)}$ is obtained by subtracting from the diagrams with internal lines in the representation $\cR$ the same diagrams with internal lines in the adjoint representation. These contributions go under the name of \emph{difference-theory} diagrams \cite{Andree:2010na,Billo:2019fbi} and, as we will see in Section \ref{sec:MM}, they tie nicely in with the corrections computed in the matrix model generated by localization on a four-sphere $S^4$.
In four-dimensional conformal theories, where the gauge coupling is a pure constant, the evanescent part $\delta^\prime	\cG^{(k)}$ is irrelevant since we can safely set $\epsilon=0$. Conversely, in theories with a non-vanishing $\beta$-function the evanescent terms interfere with the UV $\epsilon$-poles of the bare coupling 
and provide finite corrections at higher orders in perturbation theory\,\footnote{See \cite{Billo:2024fst,Billo:2024hvf} for a detailed discussion about these effects in the expectation value of  half-BPS Wilson loops.}.

For non-conformal $\mathcal{N}=2$ SQCD, the functions $\cG^{\prime\,(k)}$ were computed in \cite{Billo:2019job} for $k=1,2$, while the evanescent corrections $\delta^\prime\cG^{(k)}$ are one of the novelties of this work. In the following sections, we generalize the calculations of \cite{Billo:2019job} to any matter representation $\cR$, and show that by including the interference effects between the evanescent contributions and the UV poles of the bare coupling, the renormalized correlators perfectly match the localization predictions up to two loops.

\subsection{Tree-level}
At tree-level there is a unique diagram which can be simply computed using Wick's theorem and the scalar propagator
\begin{equation}
	\label{eq:proptree}
	\big\langle\,\phi^a(x)\,\overbar{\phi}^b(0)\,\big\rangle \,=\, 	\mathord{
		\begin{tikzpicture}[scale=0.7, baseline=-0.65ex]
			\begin{feynman}
				\vertex (A) at (-2,0) ;
				\vertex (C) at (2,0) ;
				\vertex (V) at (1.8,0.35) {\small{$0,b$} };
				\vertex (V) at (-1.8,0.35) {\small{$x,a$} };
				\diagram*{
					(A) -- [fermion] (C),
				};
			\end{feynman}
		\end{tikzpicture} 
	}  = \,\delta^{ab} \,\Delta(x,\epsilon)~.
\end{equation} 
Going through the calculation, it is straightforward to get
\begin{equation}
	\label{eq:tree-level arbitrary charge}
	\big\langle\,O_{\vec n}(x) \,\widebar{O}_{\vec m}(0)\,\big\rangle \Big|_{\mathrm{tree}}=~
	\mathord{
		\begin{tikzpicture}[scale=0.7, baseline=-0.65ex]
			\begin{feynman}
				\vertex (A) at (-2,0) ;
				\vertex (a) at (-2,0) {$\bullet$};
				\vertex (c) at (2,0) {$\bullet $ };
				\vertex (C) at (2,0) ;
				\vertex (V) at (2.6,-0.1) {\small{$0$} };
				\vertex (V) at (-2.6,-0.1) {\small{$x$} };
				\vertex (o) at (0,0) {$.$};
				\vertex (o1) at (0,0.3) {$.$};
				\vertex (o2) at (0,-0.3) {$.$};
				\diagram*{
					(A) -- [fermion, half left, looseness=0.85] (C),
					(A) --[fermion, half right, looseness=0.85] (C),
					(A) --[fermion, half right, looseness=0.55] (C),
					(A) --[fermion, half left, looseness=0.55] (C),
				};
			\end{feynman}
		\end{tikzpicture} 
	}  ~= \,g^{2n}_B \,G^{(0)}_{\vec{n},\vec{m}}\,\Delta^n(x,\epsilon) \,\delta_{nm}
\end{equation} 
where
\begin{align}
	G^{(0)}_{\vec{n},\vec{m}}=n! \,R_{\vec n}^{a_1\ldots a_n}\, R_{\vec m}^{a_1 \ldots a_n}~.
	\label{eq:cal G0 is}
\end{align}
The factor of $n!$ counts the different ways in which the $n$ indices of the two symmetric tensors $R_{\vec{n}}$ and $R_{\vec{m}}$ can be contracted among themselves. Notice that the vectors $\vec{n}$ and $\vec{m}$ can be different, but the sum of their components must be equal because of U(1) charge conservation.
Of course, in (\ref{eq:tree-level arbitrary charge}) we can safely set $\epsilon=0$ and obtain the two-point correlator in $d=4$. Here,  we give a few explicit values of $G^{(0)}_{\vec{n},\vec{m}}$ for operators of low dimensions:
\begin{subequations}
	\label{tree-explicit}
	\begin{align}
			G^{(0)}_{2,2} &= \frac{N^2-1}{2}~, \quad 
			G^{(0)}_{3,3}=\frac{3(N^2-1)(N^2-4)}{8N}~,
				\quad
			G^{(0)}_{(2,2),(2,2)} =\frac{N^4-1}{2}
				\label{G0;23}
				\\[0.4em]
			G^{(0)}_{4,(2,2)}&=\frac{(N^2-1)(2N^2-3)}{2N}~,\quad
			G^{(0)}_{4,4}=\frac{(N^2-1)(N^4-6N^2+18)}{4N^2}~.
	\end{align}
	\label{G0234}%
\end{subequations}
The expressions of $G^{(0)}_{\vec{n},\vec{m}}$ for higher dimensional operators can be calculated straightforwardly using the fusion/fission identities presented in \cite{Billo:2017glv}.

\subsection{One loop}
At one-loop accuracy ($k=1$), the two-point function \eqref{OnOm} receives contributions from the following classes of diagrams 
\begin{align}
	\label{eq:1-loop}
	&{} 
	\mathord{
\begin{tikzpicture}[scale=0.6, baseline=-0.65ex]
	\newcommand\tmpda{-0.45cm}
	\newcommand\tmpdb{-2.45cm}
	\newcommand\tmpdf{1.5cm}
	\begin{feynman}
		\vertex (A) at (-2,0) ;
		\vertex (a) at (-2,0) {$\bullet$};
		\vertex (c) at (2,0) {$\bullet $ };
		\vertex (C) at (2,0) ;
		\vertex (o1) at (0,-0.2) {$.$};
		\vertex (o2) at (0,-0.4) {$.$};
		\vertex (V) at (2.6,0) {\small{$0$} };
		\vertex (V) at (-2.6,0) {\small{$x$} };
		\vertex (i) at (-0.85,0.85);
		\vertex (j) at (0.85,0.85);
		\diagram*{
			(A) -- [plain, half left, looseness=0.85, with arrow=\tmpda] (C),
			(A) -- [plain, half left, looseness=0.85, with arrow=\tmpdb] (C),
			(A) --[fermion, half right, looseness=0.95] (C),
			(A) --[fermion, half right, looseness=0.55] (C),
		};
		\filldraw[color=white!80, fill=white!15](0,0.85) circle (0.85);	
		\diagram*{
			(j)--[charged scalar, half left,thick] (i),
			(j)--[charged scalar, half right,thick] (i),
		};
	\end{feynman}
\end{tikzpicture}
}
~,\quad
\mathord{
\begin{tikzpicture}[scale=0.6, baseline=-0.65ex]
	\newcommand\tmpda{-0.45cm}
	\newcommand\tmpdb{-2.45cm}
	\newcommand\tmpdf{1.5cm}
	\begin{feynman}
		\vertex (A) at (-2,0) ;
		\vertex (a) at (-2,0) {$\bullet$};
		\vertex (c) at (2,0) {$\bullet $ };
		\vertex (C) at (2,0) ;
		\vertex (o) at (0,0.4) {$.$};
		\vertex (o1) at (0,0.2) {$.$};
		\vertex (o2) at (0,0.0) {$.$};
		\vertex (V) at (2.6,0) {\small{$0$} };
		\vertex (V) at (-2.6,0) {\small{$x$} };
		\vertex (i) at (-1,0.85);
		\vertex (j) at (1,0.85);
		\diagram*{
			(A) -- [plain, half left, looseness=0.85, with arrow=\tmpda] (C),
			(A) -- [plain, half left, looseness=0.85, with arrow=\tmpdb] (C),
			(A) -- [plain, half left, looseness=0.85, with arrow=\tmpdf] (C),
			(A) --[fermion, half right, looseness=0.95] (C),
			(A) --[fermion, half right, looseness=0.55] (C),
			(i)--[photon, half left,thick] (j),
		};
	\end{feynman}
\end{tikzpicture}
}  
~,\quad
\mathord{
	\begin{tikzpicture}[scale=0.6, baseline=-0.65ex]
		\newcommand\tmpda{-0.45cm}
		\newcommand\tmpdb{-2.45cm}
		\begin{feynman}
			\vertex (A) at (-2,0) ;
			\vertex (a) at (-2,0) {$\bullet$};
			\vertex (c) at (2,0) {$\bullet $ };
			\vertex (C) at (2,0) ;
			\vertex (F) at (0,1.2);
			\vertex (H) at (0,0);
			\vertex (o) at (0,-0.2) {$.$};
			\vertex (o1) at (0,-0.4) {$.$};
			\vertex (o2) at (0,-0.6) {$.$};
			\vertex (V) at (2.6,0) {\small{$0$} };
			\vertex (V) at (-2.6,0) {\small{$x$} };
			\diagram*{
				(A) -- [plain, half left, looseness=1, with arrow=\tmpda] (C),
				(A) -- [plain, half left, looseness=1, with arrow=\tmpdb] (C),
				(A) --[fermion] (H),
				(H)--[fermion] (C),
				(F)-- [thick,photon] (H),
				(A) --[fermion, half right, looseness=0.95] (C),
				(A) --[fermion, half right, looseness=0.55] (C),
			};
		\end{feynman}
	\end{tikzpicture} 
	~.
}
\end{align}
Here the dashed line represents the matter hypermultiplets in the representation $\cR$, while the wavy line denotes the propagation of the $\mathcal{N}=1$ vector multiplet sitting inside the $\mathcal{N}=2$ vector multiplet. Therefore, in terms of the decomposition (\ref{W2kis}), the first diagram accounts for $\cG_{\cR}^{(1)}$, while the second and third diagram yield $\cG_{\mathrm{v.m.}}^{(1)}$.

We can conveniently combine together the first two contribution in (\ref{eq:1-loop}) and represent their sum as
\begin{equation}
	\label{eq:1-loop interactions}
	\mathord{
		\begin{tikzpicture}[scale=0.6, baseline=-0.65ex]
			\newcommand\tmpda{-0.45cm}
			\newcommand\tmpdb{-2.45cm}
			\begin{feynman}
				\vertex (A) at (-2,0) ;
				\vertex (a) at (-2,0) {$\bullet$};
				\vertex (c) at (2,0) {$\bullet $ };
				\vertex (C) at (2,0) ;
				\vertex (o) at (0,0) {$.$};
				\vertex (o1) at (0,-0.2) {$.$};
				\vertex (o2) at (0,-0.4) {$.$};
				\vertex (V) at (2.6,0) {\small{$0$} };
				\vertex (V) at (-2.6,0) {\small{$x$} };
				\diagram*{
					(A) -- [plain, half left, looseness=0.85, with arrow=\tmpda] (C),
					(A) -- [plain, half left, looseness=0.85, with arrow=\tmpdb] (C),
					(A) --[fermion, half right, looseness=0.95] (C),
					(A) --[fermion, half right, looseness=0.55] (C),
				};
				\filldraw[color=gray!80, fill=gray!15](0,0.85) circle (0.7);	
				\draw [black,thick] (0,0.85) circle [radius=0.7cm];
			\end{feynman}
			\begin{feynman}
				\vertex (o) at (0,0.8) {$v_{2,1}$};
			\end{feynman}
	\end{tikzpicture}
	}
\end{equation} 
where the grey bubble stands for the one-loop correction to the propagator of the adjoint scalar. Thus, our task is to evaluate diagrams dressed with a one-loop scalar self-energy and  with an internal vector exchange.

\paragraph{Diagrams with a one-loop correction to the propagators}
The one-loop correction to the scalar propagator is given by
\begin{equation}
	\begin{split} 
		\label{v21 def0}
		\mathord{
			\begin{tikzpicture}[baseline=-0.65ex,scale=0.52]
				\begin{feynman}
					\vertex (a) at (-1,0) ;
					\vertex (b) at (4,0)  ;
					\vertex (c) at (0.5,0) ;
					\vertex (d) at (2.5,0) ;
					\diagram*{
						(a) -- [fermion] (c),
						(c) -- [fermion] (d),
						(d) -- [fermion] (b),
					};
					\vertex (V) at (-0.7,0.45) {\small{$x,a$} };
					\vertex (V) at (3.7,0.45) {\small{$0,b$} };
				\end{feynman}
				\filldraw[color=gray!80, fill=gray!15](1.5,0) circle (1);	
				\draw [black,thick] (1.5,0) circle [radius=1cm]; 
				\begin{feynman}
					\vertex (o) at (1.5,0) {$v_{2,1}$};
				\end{feynman}
			\end{tikzpicture}
		} &~=~ 
		\mathord{
			\begin{tikzpicture}[baseline=-0.65ex,scale=0.52]
				\begin{feynman}
					\vertex (a) at (-1,0)  ;
					\vertex (b) at (4,0)  ;
					\vertex (c) at (0.5,0) ;
					\vertex (d) at (2.5,0) ;
					\vertex (e) at (1.5, 1.3) ;
					\vertex (e) at (1.5, -1.3) ;
					\diagram*{
						(a) -- [fermion] (c),
						(d) -- [charged scalar, half left,thick] (c),
						(d) -- [ charged scalar, half right,thick] (c),
						(d) -- [fermion] (b),
					};
					\vertex (V) at (-0.7,0.45) {\small{$x,a$} };
					\vertex (V) at (3.7,0.45) {\small{$0,b$} };
				\end{feynman}
			\end{tikzpicture}
		} + 	\mathord{
			\begin{tikzpicture}[baseline=-0.65ex,scale=0.52]
				\begin{feynman}
					\vertex (a) at (-1,0)  ;
					\vertex (b) at (4,0)  ;
					\vertex (c) at (0.5,0) ;
					\vertex (d) at (2.5,0) ;
					\vertex (e) at (1.5, 1.3) ;
					\vertex (e) at (1.5, -1.3) ;
					\diagram*{
						(a) -- [fermion] (c),
						(c) -- [photon, half left,thick] (d),
						(c) -- [ fermion, half right] (d),
						(d) -- [fermion] (b),
					};
					\vertex (V) at (-0.7,0.45) {\small{$x,a$} };
					\vertex (V) at (3.7,0.45) {\small{$0,b$} };
				\end{feynman}
			\end{tikzpicture}
		}~.
		\end{split}
	\end{equation} 
The last diagram receives contributions only from the $\mathcal{N}=2$ vector multiplet and thus is in common with $\mathcal{N}=4$ SYM. Exploiting the fact that in $\mathcal{N}=4$ SYM the total one-loop correction to the scalar propagator identically vanishes in \emph{any dimension}, we can write
\begin{align}
\mathord{
			\begin{tikzpicture}[baseline=-0.65ex,scale=0.52]
				\begin{feynman}
					\vertex (a) at (-1,0)  ;
					\vertex (b) at (4,0)  ;
					\vertex (c) at (0.5,0) ;
					\vertex (d) at (2.5,0) ;
					\vertex (e) at (1.5, 1.3) ;
					\vertex (e) at (1.5, -1.3) ;
					\diagram*{
						(a) -- [fermion] (c),
						(c) -- [photon, half left,thick] (d),
						(c) -- [ fermion, half right] (d),
						(d) -- [fermion] (b),
					};
					\vertex (V) at (-0.7,0.45) {\small{$x,a$} };
					\vertex (V) at (3.7,0.45) {\small{$0,b$} };
				\end{feynman}
			\end{tikzpicture}
		}
		=- \mathord{
		\begin{tikzpicture}[baseline=-0.65ex,scale=0.52]
			\begin{feynman}
				\vertex (a) at (-1,0)  ;
				\vertex (b) at (4,0)  ;
				\vertex (c) at (0.5,0) ;
				\vertex (d) at (2.5,0) ;
				\vertex (e) at (1.5, 1.3) ;
				\vertex (e) at (1.5, -1.3) ;
				\diagram*{
					(a) -- [fermion] (c),
					(d) -- [fermion, half left] (c),
					(d) -- [ fermion, half right] (c),
					(d) -- [fermion] (b),
				};
				\vertex (V) at (-0.7,0.45) {\small{$x,a$} };
				\vertex (V) at (3.7,0.45) {\small{$0,b$} };
			\end{feynman}
		\end{tikzpicture}
	} 
\end{align}
where the loop in the right-hand side is associated with the hypermultiplet in the adjoint representation. Therefore, (\ref{v21 def0}) becomes
\begin{equation}
	\begin{split} 
		\label{v21 def}
		\mathord{
			\begin{tikzpicture}[baseline=-0.65ex,scale=0.52]
				\begin{feynman}
					\vertex (a) at (-1,0) ;
					\vertex (b) at (4,0)  ;
					\vertex (c) at (0.5,0) ;
					\vertex (d) at (2.5,0) ;
					\diagram*{
						(a) -- [fermion] (c),
						(c) -- [fermion] (d),
						(d) -- [fermion] (b),
					};
					\vertex (V) at (-0.7,0.45) {\small{$x,a$} };
					\vertex (V) at (3.7,0.45) {\small{$0,b$} };
				\end{feynman}
				\filldraw[color=gray!80, fill=gray!15](1.5,0) circle (1);	
				\draw [black,thick] (1.5,0) circle [radius=1cm]; 
				\begin{feynman}
					\vertex (o) at (1.5,0) {$v_{2,1}$};
				\end{feynman}
			\end{tikzpicture}
		} &~=~ 
		\mathord{
			\begin{tikzpicture}[baseline=-0.65ex,scale=0.52]
				\begin{feynman}
					\vertex (a) at (-1,0)  ;
					\vertex (b) at (4,0)  ;
					\vertex (c) at (0.5,0) ;
					\vertex (d) at (2.5,0) ;
					\vertex (e) at (1.5, 1.3) ;
					\vertex (e) at (1.5, -1.3) ;
					\diagram*{
						(a) -- [fermion] (c),
						(d) -- [charged scalar, half left,thick] (c),
						(d) -- [ charged scalar, half right,thick] (c),
						(d) -- [fermion] (b),
					};
					\vertex (V) at (-0.7,0.45) {\small{$x,a$} };
					\vertex (V) at (3.7,0.45) {\small{$0,b$} };
				\end{feynman}
			\end{tikzpicture}
		} - 	\mathord{
			\begin{tikzpicture}[baseline=-0.65ex,scale=0.52]
				\begin{feynman}
					\vertex (a) at (-1,0)  ;
					\vertex (b) at (4,0)  ;
					\vertex (c) at (0.5,0) ;
					\vertex (d) at (2.5,0) ;
					\vertex (e) at (1.5, 1.3) ;
					\vertex (e) at (1.5, -1.3) ;
					\diagram*{
						(a) -- [fermion] (c),
						(d) -- [fermion, half left] (c),
						(d) -- [ fermion, half right] (c),
						(d) -- [fermion] (b),
					};
					\vertex (V) at (-0.7,0.45) {\small{$x,a$} };
					\vertex (V) at (3.7,0.45) {\small{$0,b$} };
				\end{feynman}
			\end{tikzpicture}
		} \\[2mm]
		&~\equiv~ 	\mathord{
			\begin{tikzpicture}[baseline=-0.65ex,scale=0.52]
				\begin{feynman}
					\vertex (a) at (-1,0) ;
					\vertex (b) at (4,0)  ;
					\vertex (c) at (0.5,0) ;
					\vertex (d) at (2.5,0) ;
					\diagram*{
						(a) -- [fermion] (c),
						(c) -- [fermion] (d),
						(d) -- [fermion] (b),
					};
					\vertex (V) at (-0.7,0.45) {\small{$x,a$} };
					\vertex (V) at (3.7,0.45) {\small{$0,b$} };
				\end{feynman}
				\filldraw[color=white!80, fill=white!15](1.5,0) circle (1);	
				\draw [black,thick] (1.5,0) circle [radius=1cm]; 
				\draw [black,thick,dashed] (1.5,0) circle [radius=0.9cm];
				\end{tikzpicture}
			} \,= \, g_B^2\,v_{2,1}(x,\epsilon)\,\Delta(x,\epsilon)\,\delta_{ab} \ .
		\end{split}
	\end{equation} 
Here we have introduced the double dashed/continuous line to emphasize that this combination of diagrams is a difference-theory contribution and defined\,\footnote{We evaluate the diagrams using the $\mathcal{N}=1$ superfield formalism. In this approach, the scalar $\phi(x)$ corresponds to the lowest component of the $\mathcal{N}=1$ chiral superfield $\Phi(x,\theta,\bar{\theta})$ sitting inside the $\mathcal{N}=2$ vector multiplet. Our conventions for the Feynman rules in this super-space formalism are the same as in \cite{Billo:2017glv,Billo:2019job} to which we refer for details.}
	\begin{equation}
		v_{2,1}(x,\epsilon) =\beta_0 \,\frac{(\pi x^2)^{\epsilon}}{8\pi^2}
		\frac{\Gamma(1-\epsilon)}{2\epsilon(1-2\epsilon)}
		\label{v21}
	\end{equation} 
where $\beta_0$ is the $\beta$-function coefficient (\ref{beta0gen}). This function exhibits a UV divergence, associated with the pole in $\epsilon$, and vanishes in superconformal models where $\beta_0=0$.  

Using this result, it is straightforward to calculate the contribution of the diagram in  (\ref{eq:1-loop interactions}) and get
	\begin{equation}
		\begin{split}
			\label{eq:insertion of the one-loop correction}
			\mathord{
				\begin{tikzpicture}[scale=0.6, baseline=-0.65ex]
					\newcommand\tmpda{-0.45cm}
					\newcommand\tmpdb{-2.45cm}
					\begin{feynman}
						\vertex (A) at (-2,0) ;
						\vertex (a) at (-2,0) {$\bullet$};
						\vertex (c) at (2,0) {$\bullet $ };
						\vertex (C) at (2,0) ;
						\vertex (o) at (0,0) {$.$};
						\vertex (o1) at (0,-0.2) {$.$};
						\vertex (o2) at (0,-0.4) {$.$};
						\vertex (V) at (2.6,0) {\small{$0$} };
						\vertex (V) at (-2.6,0) {\small{$x$} };
						\diagram*{
							(A) -- [plain, half left, looseness=0.85, with arrow=\tmpda] (C),
							(A) -- [plain, half left, looseness=0.85, with arrow=\tmpdb] (C),
							(A) --[fermion, half right, looseness=0.95] (C),
							(A) --[fermion, half right, looseness=0.55] (C),
						};
						\filldraw[color=gray!80, fill=gray!15](0,0.85) circle (0.7);	
						\draw [black,thick] (0,0.85) circle [radius=0.7cm];
					\end{feynman}
					\begin{feynman}
						\vertex (o) at (0,0.8) {$v_{2,1}$};
					\end{feynman}
			\end{tikzpicture}}\,=\,g_B^{2n+2} \, n\, v_{2,1}(x,\epsilon)\,G^{(0)}_{\vec{n},\vec{m}}\, \Delta^n(x,\epsilon) ~ 
		\end{split}
	\end{equation} 
where $n$ is the  symmetry factor corresponding to the insertion of the bubble in each of the $n$ propagators. According to the decomposition \eqref{calG def}, we have
\begin{align}
\label{G2prime}
\cG^{\prime\,(1)}_{\vec{n},\vec{m}}(x,\epsilon)= n\, v_{2,1}(x,\epsilon)~
\end{align} since (\ref{eq:insertion of the one-loop correction})  is a pure difference-theory contribution.
	
\paragraph{Diagrams with internal vector lines}
To compute the contribution of the rightmost diagram in 
\eqref{eq:1-loop}, it is convenient to first
consider the following building block diagram (see Appendix \refeq{sec:pertcalc} for some details):
	\begin{equation}
		\label{v42Aform}
		\mathord{
			\begin{tikzpicture}[scale=0.5, baseline=-0.65ex]	
				\begin{feynman}
					\vertex (A) at (0,1.2);
					\vertex (B) at (0,-1.2);
					\vertex (a) at (-4.1,2) {\small{$x, a_1$}};
					\vertex (a1) at (-4,1.2) ;
					\vertex (b1) at (-1.1,1.2) ;
					\vertex (d) at (-4.1,-2.) {\small{$x, a_2$}};
					\vertex (d1) at (-4,-1.2) ;
					\vertex (c1) at (-1.1,-1.2) ;
					\vertex (e) at (4.1,2) {\small{$0, b_1$}};
					\vertex (e1) at (4,1.2) ;
					\vertex (f1) at (1.1,1.2) ;
					\vertex (h) at (4.1,-2.) {\small{$0, b_2$}};
					\vertex (h1) at (4,-1.2) ;
					\vertex (g1) at (1.1,-1.2) ;
					\diagram*{
						(b1) -- [plain] (e1),
						(c1) -- [plain] (h1),
						(a1)--[fermion] (b1),
						(d1)--[fermion] (c1),
						(f1)--[fermion] (e1),
						(g1)--[fermion] (h1),
						(A)--[thick,photon] (B),
					};
				\end{feynman}
			\end{tikzpicture} 
		} =  g_B^2 \,
		C^{a_1 a_2 b_1 b_2} 
		\,e_{2,1}(x,\epsilon)\,\Delta^2(x,\epsilon)
	\end{equation} 
where the tensor 
	\begin{equation}
		\label{eq:C4A}
		C^{a_1a_2 b_1 b_2} = -\frac{1}{N}\,f^{a_1 b_1 c}\,f^{a_2 b_2 c}
	\end{equation}
encodes the color structure, and the function
	\begin{equation}
		\label{eq:e21is}
		e_{2,1}(x,\epsilon) = \frac{3 N \,\zeta (3) \,\epsilon \,(x^2)^{\epsilon } \,\Gamma (2-3 \epsilon )}{4^{1+2 \epsilon} \,\pi ^{2+\epsilon} \,\Gamma (1-\epsilon )^2\, \Gamma (1+2\epsilon)}=
		\epsilon \,\frac{3N\,\zeta(3)}{4\pi^2}   +\cO(\epsilon^2)
	\end{equation}
captures the space-time dependence of the diagram.
Note that this term is \emph{evanescent}  when $d \to 4$.
	
Using \eqref{v42Aform} and exploiting the fusion/fission rules of SU$(N)$ to compute the color factor of the corresponding correlator as explained in \cite{Billo:2019job}\,\footnote{See in particular Eq.s (2.36)-(2.39) of \cite{Billo:2019job}.}, we obtain
	\begin{equation}
		\label{eq:internal exchange def}
		\mathord{
			\begin{tikzpicture}[scale=0.6, baseline=-0.65ex]
				\newcommand\tmpda{-0.45cm}
				\newcommand\tmpdb{-2.45cm}
				\begin{feynman}
					\vertex (A) at (-2,0) ;
					\vertex (a) at (-2,0) {$\bullet$};
					\vertex (c) at (2,0) {$\bullet $ };
					\vertex (C) at (2,0) ;
					\vertex (F) at (0,1.2);
					\vertex (H) at (0,0);
					\vertex (o) at (0,-0.2) {$.$};
					\vertex (o1) at (0,-0.4) {$.$};
					\vertex (o2) at (0,-0.6) {$.$};
					\vertex (V) at (2.6,0) {\small{$0$} };
					\vertex (V) at (-2.6,0) {\small{$x$} };
					\diagram*{
						(A) -- [plain, half left, looseness=1, with arrow=\tmpda] (C),
						(A) -- [plain, half left, looseness=1, with arrow=\tmpdb] (C),
						(A) --[fermion] (H),
						(H)--[fermion] (C),
						(F)-- [thick,photon] (H),
						(A) --[fermion, half right, looseness=0.95] (C),
						(A) --[fermion, half right, looseness=0.55] (C),
					};
				\end{feynman}
			\end{tikzpicture} 
		}  =- g_B^{2n+2}\,n\,e_{2,1}(x,\epsilon)\,G^{(0)}_{\vec{n},\vec{m}}\, \Delta^n(x,\epsilon) ~.
	\end{equation} 
This is an explicit realization of an evanescent contribution and thus, according to \eqref{calG def}, we can write
\begin{align}
\label{deltaprimeG2}
\delta^\prime\cG^{(1)}_{\vec{n},\vec{m}}(x,\epsilon)=- n\, e_{2,1}(x,\epsilon)~.
\end{align}
Combining together (\ref{G2prime}) and (\ref{deltaprimeG2}), we find that the total one-loop result is
	\begin{equation}
				\cG^{(1)}_{\vec{n},\vec{m}}(x,\epsilon)= \cG^{\prime\,(1)}_{\vec{n},\vec{m}}(x,\epsilon)+ \delta^\prime \cG^{(1)}_{\vec{n},\vec{m}}(x,\epsilon)= n \big[v_{2,1}(x,\epsilon)-e_{2,1}(x,\epsilon)\big]~.
		\label{G1loop}
	\end{equation} 
The previous expression confirms that, up to evanescent terms, the one-loop corrections are encoded in the difference-theory term $\cG^{\prime\,(1)}_{\vec{n},\vec{m}}$ given in (\ref{G2prime}). At two loops, we exploit this fact to reduce the number of diagrams we have to compute. 
\subsection{Two loops}
We now consider the two-loop contributions ($k=2$). Since we are interested 
in computing correlators up to order $g_B^{2n+4}$, we can focus on the difference-theory terms and discard  the two-loop evanescent parts which only affect the result at order $g_B^{2n+6}$.
Most of the difference-theory diagrams at this order were already computed in \cite{Billo:2019job} for the non-conformal $\mathcal{N}=2$ SQCD. Here, we generalize these results to a generic representation $\cR$ and also  include contributions containing the one-loop evanescent factor $e_{2,1}(x,\epsilon)$ that were not considered in \cite{Billo:2019job}. 

\paragraph{Diagrams with two one-loop corrections to the propagators}
At two loops the correlator (\ref{OnOm}) receives contributions from diagrams in which two propagators are dressed with the self-energy \eqref{v21 def}, namely
	\begin{equation}
		\mathord{
			\begin{tikzpicture}[scale=0.8, baseline=-0.65ex]
				\newcommand\tmpda{-0.85cm}
				\newcommand\tmpdb{-3.25cm}
				\begin{feynman}
					\vertex (A) at (-2,0) ;
					\vertex (a) at (-2,0) {$\bullet$};
					\vertex (c) at (2,0) {$\bullet $ };
					\vertex (C) at (2,0) ;
					\vertex (V) at (2.6,-0.1) {$0$ };
					\vertex (v) at (-2.6,-0.1) {$x$ };
					\vertex (o1) at (0,0.1) {$\vdots$};
					\diagram*{
						(A) -- [plain, half right, looseness=1.1,with arrow=\tmpda] (C),
						(A) -- [plain, half right, looseness=1.1,with arrow=\tmpdb] (C),
						(A) -- [plain, half left, looseness=1.1,with arrow=\tmpdb] (C),
						(A) -- [plain, half left, looseness=1.1,with arrow=\tmpda] (C),
						(A) -- [fermion, half left, looseness=0.45] (C),
						(A) -- [fermion, half right, looseness=0.45] (C),
					};
					\filldraw[color=gray!80, fill=gray!15](0,1.25) circle (0.5);	
					\draw [black,thick] (0,1.25) circle [radius=0.5cm];
					\filldraw[color=gray!80, fill=gray!15](0,-1.25) circle (0.5);
					\draw [black,thick] (0,-1.25) circle [radius=0.5cm];
				\end{feynman}
				\begin{feynman}
					\vertex (o) at (0,1.25) {\small{$v_{2,1}$}};
				\end{feynman}
				\begin{feynman}
					\vertex (o) at (0,-1.25) {\small{$v_{2,1}$}};
				\end{feynman}
			\end{tikzpicture} 
		} = g_B^{2n+4}\,\frac{n(n-1)}{2} \,v_{2,1}^2(x,\epsilon)\,G^{(0)}_{\vec{n},\vec{m}}\,\Delta^n(x,\epsilon)
	\end{equation}
where the symmetry factor $\frac{n(n-1)}{2}$ takes into account the combinatorics of the possible bubble insertions. 

Similarly, we can consider a single scalar propagator with two one-loop corrections \eqref{v21 def} and get
	\begin{equation}
		\mathord{
			\begin{tikzpicture}[scale=0.8, baseline=-0.65ex]
				\newcommand\tmpda{-0.45cm}
				\newcommand\tmpdb{-2.85cm}
				\newcommand\tmpdc{-3.65cm}
				\newcommand\tmpdd{-.55cm}
				\begin{feynman}
					\vertex (A) at (-2,0) ;
					\vertex (a) at (-2,0) {$\bullet$};
					\vertex (c) at (2,0) {$\bullet $ };
					\vertex (C) at (2,0) ;
					\vertex (V) at (2.6,-0.1) {$0$ };
					\vertex (v) at (-2.6,-0.1) {$x$ };
					\vertex (o1) at (0,0.1) {$\vdots$};
					\diagram*{
						(A) -- [fermion, half left, looseness=1.1] (C),
						(A) -- [fermion, half right, looseness=1.1] (C),
						(A) -- [plain, half left, looseness=1.1,with arrow=\tmpdc] (C),
						(A) -- [plain, half left, looseness=1.1,with arrow=\tmpdd] (C),
						(A) -- [fermion, half left, looseness=0.45] (C),
						(A) -- [fermion, half right, looseness=0.45] (C),
					};
					\filldraw[color=gray!80, fill=gray!15](-1,1.1) circle (0.5);	
					\draw [black,thick] (-1,1.1) circle [radius=0.5cm];
					\filldraw[color=gray!80, fill=gray!15](1,1.1) circle (0.5);
					\draw [black,thick] (1,1.1) circle [radius=0.5cm];
				\end{feynman}
				\begin{feynman}
					\vertex (o) at (1,1.1) {\small{$v_{2,1}$}};
				\end{feynman}
				\begin{feynman}
					\vertex (o) at (-1,1.1) {\small{$v_{2,1}$}};
				\end{feynman}
			\end{tikzpicture} 
		} = g_B^{2n+4}\,n \,v_{2,1}^2(x,\epsilon)\,G^{(0)}_{\vec{n},\vec{m}}\,\Delta^n(x,\epsilon) 
	\end{equation}
where $n$ is the combinatorial factor of the diagram.
	
\paragraph{Diagrams with a two-loop correction to the propagators}
The two-loop correction to the scalar propagator is described by the following diagrams evaluated in \cite{Billo:2019job}:
	\begin{equation}
		\label{v22 def}
		\begin{split}
			\mathord{
				\begin{tikzpicture}[baseline=-0.65ex,scale=0.52]
					\begin{feynman}
						\vertex (a) at (-1,0) ;
						\vertex (b) at (4,0)  ;
						\vertex (c) at (0.5,0) ;
						\vertex (d) at (2.5,0) ;
						\diagram*{
							(a) -- [fermion] (c),
							(c) -- [fermion] (d),
							(d) -- [fermion] (b),
						};
						\vertex (V) at (-0.7,0.45) {\small{$x,a$} };
						\vertex (V) at (3.7,0.45) {\small{$0,b$} };
					\end{feynman}
					\filldraw[color=gray!80, fill=gray!15](1.5,0) circle (1);	
					\draw [black,thick] (1.5,0) circle [radius=1cm]; 
					\begin{feynman}
						\vertex (o) at (1.5,0) {$v_{2,2}$};
					\end{feynman}
				\end{tikzpicture}
			} 
			& =\, \mathord{
				\begin{tikzpicture}[scale=0.5, baseline=-0.65ex]
					\filldraw[color=white!80, fill=white!15](0,0) circle (0.8);	
					\draw [black] (0,0) circle [radius=0.8cm];
					\draw [black, thick, dashed] (0,0) circle [radius=0.7cm];
					\begin{feynman}
						\vertex (a) at (-2.8,0);
						\vertex (A) at (-1.8,0);
						\vertex (C) at (-0.8,0);
						\vertex (B) at (0.8, 0);
						\vertex (D) at (1.8, 0);
						\vertex (d) at (2.8,0);
						\diagram*{
							(a) -- [ fermion] (A),
							(A)-- [fermion] (C),
							(B) --[fermion] (D),
							(A) --[photon, half left]  (D),
							(D) --[fermion] (d)
						};
						\vertex (V) at (-2.5,0.45) {\small{$x,a$} };
						\vertex (V) at (2.7,0.45) {\small{$0,b$} };
					\end{feynman}
				\end{tikzpicture} 
			} +
			\mathord{
				\begin{tikzpicture}[scale=0.5, baseline=-0.65ex]
					\begin{feynman}
						\vertex (a) at (-2.8,0);
						\vertex (A) at (-1.8,0);
						\vertex (C) at (-0.8,0);
						\vertex (B) at (0.8, 0);
						\vertex (b1) at (-0.8,1.5);
						\vertex (b2) at (0.8,1.5);
						\vertex (D) at (1.8, 0);
						\vertex (d) at (2.8,0);
						\diagram*{
							(a) -- [fermion] (A),
							(A) -- [fermion] (D),
							(A) --[photon, half left]  (D),
							(D)-- [fermion] (d),
						};
						\filldraw[color=white, fill=white](0,1.5) circle (0.8);	
						\draw [black] (0,1.5) circle [radius=0.8cm];
						\draw [black, thick, dashed] (0,1.5) circle [radius=0.7cm];
						\vertex (V) at (-2.5,0.45) {\small{$x,a$} };
						\vertex (V) at (2.7,0.45) {\small{$0,b$} };
					\end{feynman}
				\end{tikzpicture} 
			} \\
			& ~\quad+ 	\mathord{
				\begin{tikzpicture}[baseline=-0.6ex,scale=0.5]
					\draw[
					]
					(1.5,0) circle (1);
					\draw[
					thick, black,dashed
					]
					(1.5,0) circle (0.9);
					\begin{feynman}
						\vertex (a) at (-1,0)  ;
						\vertex (c) at (0.5,0) ;
						\vertex (d) at (2.5,0) ;
						\vertex (d1) at (1,1) ;
						\vertex (A) at (1.5,1);
						\vertex (B) at (1.5,-1);
						\diagram*{
							(a) -- [fermion](c),
							(d) -- [fermion] (b),
							(A) --[photon] (B)
						};
						\vertex (V) at (-0.7,0.45) {\small{$x,a$} };
						\vertex (V) at (4.0,0.45) {\small{$0,b$} };
					\end{feynman}
				\end{tikzpicture}
			} + 	\mathord{\begin{tikzpicture}[baseline=-0.65ex,scale=0.5]  
					\draw[
					]
					(1.5,0) circle (1);
					\draw[
					thick, black,dashed
					]
					(1.5,0) circle (0.9);
					\begin{feynman}
						\vertex (a) at (-1,0)  ;
						\vertex (b) at (4.2,0) ;
						\vertex (c) at (0.5,0) ;
						\vertex (c1) at (1.5,-0.9);
						\vertex (c3) at (3.25,0);
						\vertex (c2) at (0.4, 0);
						\vertex (d1) at (1.5,1);
						\vertex (d4) at (1.5,-0.9);
						\vertex (d) at (2.5,0) ;
						\vertex (d2) at (2.6,0);
						\vertex (e) at (1.5, 1.5) ;
						\vertex (e) at (1.5, -1.5) ;
						\diagram*{
							(a) -- [fermion] (c),
							(d) -- [fermion] (c3),
							(d1)--[photon, half left] (c3),
							(c3) --[fermion] (b),
						};
						\vertex (V) at (-0.7,0.45) {\small{$x,a$} };
						\vertex (V) at (4.15,0.45) {\small{$0,b$} };
					\end{feynman}
			\end{tikzpicture}	} \\[4mm]
			&
			=\,g_B^4\,v_{2,2}(x,\epsilon)\,\Delta(x,\epsilon)\,\delta^{ab}
		\end{split}
	\end{equation}
where
	\begin{align}
		v_{2,2}(x,\epsilon)=\frac{(\pi x^2)^{2\epsilon}}{(8\pi^2)^2 }\bigg[6\,\zeta(3)\Big(C_\cR\,i_\cR-N^2+\frac{N\beta_0}{4}\Big)+ \frac{N\beta_0\,\Gamma^2(1-\epsilon)}{4\epsilon^2\,(1-2\epsilon)(1+\epsilon)}\bigg] +\mathcal{O}(\epsilon)
		\label{v22is}
	\end{align}
with $C_\cR$ denoting the quadratic Casimir invariant of the representation $\cR$. For the special representation $\cR$ (\ref{eq:Ris}), the overall coefficient of $\zeta(3)$ inside the square brackets becomes
\begin{align}
-3\bigg(\frac{N_f}{2N}+\frac{N_a(N-2)(4+2N-N^2)}{2N}+\frac{N_s(N+2)(4-2N-N^2)}{2N}+N^2\bigg)
\label{K22a}
\end{align}
which reduces to the one given in \cite{Billo:2019job} for SQCD. Notice that the term in (\ref{v22is}) proportional to $\zeta(3)$ is regular when $\epsilon\to 0$, while the second term proportional to $N\beta_0$ exhibits a double pole in $\epsilon$.
Using this result, it is immediate to find that 
	\begin{equation}
		\begin{split}
			\label{eq:insertion of the two-loop correction}
			\mathord{
				\begin{tikzpicture}[scale=0.6, baseline=-0.65ex]
					\newcommand\tmpda{-0.45cm}
					\newcommand\tmpdb{-2.45cm}
					\begin{feynman}
						\vertex (A) at (-2,0) ;
						\vertex (a) at (-2,0) {$\bullet$};
						\vertex (c) at (2,0) {$\bullet $ };
						\vertex (C) at (2,0) ;
						\vertex (o) at (0,0) {$.$};
						\vertex (o1) at (0,-0.2) {$.$};
						\vertex (o2) at (0,-0.4) {$.$};
						\vertex (V) at (2.6,0) {\small{$0$} };
						\vertex (V) at (-2.6,0) {\small{$x$} };
						\diagram*{
							(A) -- [plain, half left, looseness=0.85, with arrow=\tmpda] (C),
							(A) -- [plain, half left, looseness=0.85, with arrow=\tmpdb] (C),
							(A) --[fermion, half right, looseness=0.95] (C),
							(A) --[fermion, half right, looseness=0.55] (C),
						};
						\filldraw[color=gray!80, fill=gray!15](0,0.85) circle (0.7);	
						\draw [black,thick] (0,0.85) circle [radius=0.7cm];
					\end{feynman}
					\begin{feynman}
						\vertex (o) at (0,0.8) {$v_{2,2}$};
					\end{feynman}
			\end{tikzpicture}}\,=\,g_B^{2n+4} \, n\, v_{2,2}(x,\epsilon)\,G^{(0)}_{\vec{n},\vec{m}}\, \Delta^n(x,\epsilon) 
		\end{split}
	\end{equation} 
where $n$ is the symmetry factor of the diagram.
	
\paragraph{Diagrams with internal vector lines and one-loop self-energies}
At two loops,  the correlator \eqref{OnOm} receives contributions also from diagrams involving vector-multiplet propagation and a single scalar self-energy (\ref{v21 def}). These contributions can be organized in two distinct families of corrections according to the insertion of the self-energy. 
For instance, when dressing with the one-loop correction (\ref{v21 def}) the lines of the diagram (\ref{v42Aform}), we obtain the diagrams originally analyzed in 
\cite{Billo:2019job}, namely
		\begin{equation}
		\begin{split}
			\mathord{
				\begin{tikzpicture}[scale=0.5, baseline=-0.65ex]	
					\begin{feynman}
						\vertex (A) at (0,2);
						\vertex (B) at (0,-2);
						\vertex (a) at (-3,2.3) {\small{$x, a_1$}};
						\vertex (a1) at (-3.,1.5) ;
						\vertex (b1) at (-0.7,1.5) ;
						\vertex (d) at (-3,-2.3) {\small{$x, a_2$}};
						\vertex (d1) at (-3.,-1.5) ;
						\vertex (c1) at (-0.7,-1.5) ;
						\vertex (e) at (3,2.3) {\small{$0, b_1$}};
						\vertex (e1) at (3.,1.5) ;
						\vertex (f1) at (0.7,1.5) ;
						\vertex (h) at (3,-2.3) {\small{$0, b_2$}};
						\vertex (h1) at (3.,-1.5) ;
						\vertex (g1) at (0.7,-1.5) ;
						\diagram*{
							(a1)--[fermion] (b1),
							(d1)--[fermion] (c1),
							(f1)--[fermion] (e1),
							(g1)--[fermion] (h1),
						};
					\end{feynman}
					\filldraw[color=black,thick, fill=gray!15](0,0) ellipse (1.1 and 2.1);
					\begin{feynman}
						\vertex (O) at (0,0){$v_{4,2}$} ;
					\end{feynman}
				\end{tikzpicture} 
			} &\equiv 		~	\mathord{
				\begin{tikzpicture}[scale=0.5, baseline=-0.65ex]	
					\begin{feynman}
						\vertex (A) at (0,1.5);
						\vertex (B) at (0,-1.5);
						\vertex (a) at (-3,2.3) {\small{$x, a_1$}};
						\vertex (a1) at (-3.,1.5) ;
						\vertex (b1) at (-0.1,1.5) ;
						\vertex (d) at (-3,-2.3) {\small{$x, a_2$}};
						\vertex (d1) at (-3.,-1.5) ;
						\vertex (c1) at (-0.1,-1.5) ;
						\vertex (e) at (3,2.3) {\small{$0, b_1$}};
						\vertex (e1) at (3.,1.5) ;
						\vertex (f1) at (0.1,1.5) ;
						\vertex (h) at (3,-2.3) {\small{$0, b_2$}};
						\vertex (h1) at (3.,-1.5) ;
						\vertex (g1) at (0.1,-1.5) ;
						\diagram*{
							(b1) -- [plain] (e1),
							(c1) -- [plain] (h1),
							(a1)--[fermion] (b1),
							(d1)--[fermion] (c1),
							(f1)--[fermion] (e1),
							(g1)--[fermion] (h1),
							(A)--[photon] (B),
						};
						\filldraw[color=white!80, fill=white!15](0,-1.3) circle (0.75);	
						\draw [black,thick] (0,-1.5) circle [radius=0.85cm];				
						\draw [black,thick,dashed] (0,-1.5) circle [radius=0.73cm];
					\end{feynman}
				\end{tikzpicture} 
			} 
			+ 
			\mathord{
				\begin{tikzpicture}[scale=0.5, baseline=-0.65ex]	
					\begin{feynman}
						\vertex (A) at (0,1.5);
						\vertex (B) at (0,-1.5);
						\vertex (a) at (-3,2.3) {\small{$x, a_1$}};
						\vertex (a1) at (-3.,1.5) ;
						\vertex (b1) at (-0.1,1.5) ;
						\vertex (d) at (-3,-2.3) {\small{$x, a_2$}};
						\vertex (d1) at (-3.,-1.5) ;
						\vertex (c1) at (-0.1,-1.5) ;
						\vertex (e) at (3,2.3) {\small{$0, b_1$}};
						\vertex (e1) at (3.,1.5) ;
						\vertex (f1) at (0.1,1.5) ;
						\vertex (h) at (3,-2.3) {\small{$0, b_2$}};
						\vertex (h1) at (3.,-1.5) ;
						\vertex (g1) at (0.1,-1.5) ;
						\diagram*{
							(b1) -- [plain] (e1),
							(c1) -- [plain] (h1),
							(a1)--[fermion] (b1),
							(d1)--[fermion] (c1),
							(f1)--[fermion] (e1),
							(g1)--[fermion] (h1),
							(A)--[photon] (B),
						};
						\filldraw[color=white!80, fill=white!15](0,0) circle (0.81);	
						\draw [black,thick] (0,0) circle [radius=0.85cm];				
						\draw [black,thick,dashed] (0,0) circle [radius=0.73cm];
					\end{feynman}
				\end{tikzpicture} 
			}\\[2mm]
			&~+
			\mathord{
				\begin{tikzpicture}[scale=0.5, baseline=-0.65ex]	
					\newcommand\tmpda{-1.25cm}
					\newcommand\tmpdb{-2.35cm}
					\newcommand\tmpdc{-1.3cm}
					\begin{feynman}
						\vertex (A) at (0.7,1.5);
						\vertex (B) at (0.7,-1.5);
						\vertex (a) at (-3.2,2.3) {\small{$x, a_1$}};
						\vertex (a1) at (-3.2,1.5) ;
						\vertex (b1) at (0.7,1.5) ;
						\vertex (d) at (-3.2,-2.3) {\small{$x, a_2$}};
						\vertex (d1) at (-3.2,-1.5) ;
						\vertex (c1) at (-0.1,-1.5) ;
						\vertex (e) at (2.8,2.3) {\small{$0, b_1$}};
						\vertex (e1) at (2.8,1.5) ;
						\vertex (f1) at (0.7,1.5) ;
						\vertex (h) at (2.8,-2.3) {\small{$0, b_2$}};
						\vertex (h1) at (2.8,-1.5) ;
						\vertex (g1) at (0.7,-1.5) ;
						\diagram*{
							(b1) -- [plain] (e1),
							(c1) -- [plain, with arrow=\tmpdc] (h1),
							(a1)--[fermion] (b1),
							(d1)--[plain, with arrow=\tmpda] (c1),
							(f1)--[fermion] (e1),
							(g1)--[fermion] (h1),
							(A)--[photon] (B),
						};
						\filldraw[color=white!80, fill=white!15](-1.3,-1.5) circle (0.81);	
						\draw [black,thick] (-1.3,-1.5) circle [radius=0.85cm];				
						\draw [black,thick,dashed] (-1.3,-1.5) circle [radius=0.73cm];
					\end{feynman}
				\end{tikzpicture} 
			} 
			\!\!\!= g_B^4 \,
			C^{a_1 a_2 b_1 b_2} 
			\,v_{4,2}(x,\epsilon)\,\Delta^2(x,\epsilon)+\cdots ~.
		\end{split}
		\label{diagrams42A}
\end{equation}
Here, the color tensor is defined in \eqref{eq:C4A} and the function $v_{4,2}(x,\epsilon)$ is given 
by\,\footnote{In \cite{Billo:2019job} this function is denoted  by $v^{(A)}_{4,2}(x,\epsilon)$ and contains also a factor of $g_B^4$ which instead here we have explicitly put in front of the whole amplitude.}
	\begin{align}
		v_{4,2}(x,\epsilon)=
		\frac{(\pi x^2)^{2\epsilon}}{(8\pi^2)^2 }\,N\beta_0\bigg[\frac{21\,\zeta(3)}{2}+ \frac{\Gamma^2(1-\epsilon)}{4\epsilon^2\,(1-2\epsilon)(1+\epsilon)}\bigg] +\mathcal{O}(\epsilon)~.
		\label{v42Ais}
	\end{align}
Finally, the ellipses on right-hand side of \eqref{diagrams42A} stand for terms that are anti-symmetric in $(a_1,a_2)$ and $(b_1,b_2)$. Since they do not contribute when this sub-diagram is inserted in a chiral/anti-chiral correlator, they can be neglected. Using this result and the fusion/fission identities of SU($N$) to compute the overall color factor, we find
	\begin{equation}
		\label{v42Acontribution}
		\begin{split}
			\mathord{
				\begin{tikzpicture}[scale=0.35, baseline=-0.65ex]	
					\newcommand\tmpda{-0.97cm}
					\newcommand\tmpdb{-2.9cm}
					\newcommand\tmpdf{-0.75cm}
					\newcommand\tmpdg{-2.45cm}
					\begin{feynman}
						\vertex (i) at (-4.,0) {$\bullet$};
						\vertex (j) at (4.,0) {$\bullet $ };
						\vertex (a) at (-4.8,0.) {\small{$x$}};
						\vertex (a1) at (-4,0) ;
						\vertex (b1) at (-0.9,0.5) ;
						\vertex (d1) at (0.8,0) ;
						\vertex (c1) at (-1.1,-1.2) ;
						\vertex (e) at (4.8,0.) {\small{$0$}};
						\vertex (e1) at (4,0) ;
						\vertex (f1) at (1.1,1.2) ;
						\vertex (h1) at (4,0) ;
						\vertex (g1) at (1.1,-1.2) ;
						\vertex (o2) at (0,-0.6) {$\vdots$};
						\diagram*{
							(a1) -- [plain, half left, looseness=1.2, with arrow=\tmpda] (h1),
							(a1) -- [plain, half left, looseness=1.2, with arrow=\tmpdb] (h1),
							(a1) -- [plain, half left, looseness=0.65, with arrow=\tmpdf] (h1),
							(a1) -- [plain, half left, looseness=0.65, with arrow=\tmpdg] (h1),
							(a1) -- [fermion, half right, looseness=1.2] (h1),
							(a1) -- [fermion, half right, looseness=0.8] (h1),
						};
					\end{feynman}
					\filldraw[color=black,thick, fill=gray!15](0,2) ellipse (1.1 and 2.1);
					\begin{feynman}
						\vertex (O) at (0,2){$v_{4,2}$} ;
					\end{feynman}
				\end{tikzpicture} 
			} \,=\,-g_B^{2n+4} \, n\, v_{4,2}(x,\epsilon)\,G^{(0)}_{\vec{n},\vec{m}}\, \Delta^n(x,\epsilon)
		\end{split}
	\end{equation}
where the sign and the factor of $n$ have the same origin of those in (\ref{eq:internal exchange def}).
	
A second class of contributions arises from diagrams where the one-loop correction is inserted in one of the scalar propagators that are not involved in the vector exchange, {\it{i.e.}}
	\begin{equation}
		\mathord{
			\begin{tikzpicture}[scale=0.5, baseline=-0.65ex]
				\newcommand\tmpda{-0.80cm}
				\newcommand\tmpdb{-2.6cm}
				\newcommand\tmpdc{-0.65cm}
				\newcommand\tmpdd{-2.20cm}
				\newcommand\tmpde{-0.95cm}
				\newcommand\tmpdf{-3.15cm}
				\begin{feynman}
					\vertex (A) at (-3,0) ;
					\vertex (a) at (-3,0) {$\bullet$};
					\vertex (c) at (3,0) {$\bullet $ };
					\vertex (C) at (3,0) ;
					\vertex (F) at (0,1.65);
					\vertex (H) at (0,0.);
					\vertex (V) at (3.9,0) {\small{$0$} };
					\vertex (V) at (-3.9,0) {\small{$x$} };
					\vertex (S) at (0,-0.4) {$\vdots$};
					\diagram*{
						(A) -- [plain, half left, looseness=0.95, with arrow=\tmpda] (C),
						(A) -- [plain, half left, looseness=0.95, with arrow=\tmpdb] (C),
						(F)-- [thick,photon] (H),
						(A) --[plain, half left, looseness=0., with arrow=\tmpdc] (C),
						(A) --[plain, half left, looseness=0., with arrow=\tmpdd] (C),
						(A) -- [fermion, half right, looseness=0.7] (C),
						(A) -- [plain, half right, looseness=1.3, with arrow=\tmpde] (C),
						(A) -- [plain, half right, looseness=1.3, with arrow=\tmpdf] (C),
					};
					\filldraw[color=black,thick, fill=gray!15](0,-2.3) circle (0.8);
				\end{feynman}
				\begin{feynman}
					\vertex (o) at (0,-2.3) {$v_{2,1}$};
				\end{feynman}
			\end{tikzpicture} 
		} =- g_B^{2n+4}\,n(n-2)\, v_{2,1}(x,\epsilon)\,e_{2,1}(x,\epsilon)\,G^{(0)}_{\vec{n},\vec{m}}\, \Delta^n(x,\epsilon)~.	
		\label{v21-e21}
	\end{equation} 
The overall symmetry factor can be explained as follows. The sign and the factor of $n$ are analogous to those appearing in the vector-exchange diagram at one loop (see \eqref{eq:internal exchange def}), while the factor of $(n-2)$ arises from the possibilities of inserting the one-loop self-energy in each of the $(n-2)$ remaining propagators. Importantly, the function appearing in \eqref{v21-e21} is finite in limit $\epsilon\to 0$, since the pole of $v_{2,1}(x,\epsilon)$ is compensated by the evanescent term $e_{2,1}(x,\epsilon)$. Therefore this contribution is regular at this order.

\paragraph{Diagram with a scalar box} 
The last two-loop correction to the correlator \eqref{OnOm} we have to consider corresponds to diagrams with an internal loop of hypermultiplets in the difference theory  \cite{Billo:2017glv,Billo:2019job}. The basic building block is   
	\begin{equation}
		\begin{split}
			\mathord{
				\begin{tikzpicture}[scale=0.5, baseline=-0.65ex]	
					\begin{feynman}
						\vertex (A) at (0,2);
						\vertex (B) at (0,-2);
						\vertex (a) at (-3,2.3) {\small{$x, a_1$}};
						\vertex (a1) at (-3.,1.5) ;
						\vertex (b1) at (-0.7,1.5) ;
						\vertex (d) at (-3,-2.3) {\small{$x, a_2$}};
						\vertex (d1) at (-3.,-1.5) ;
						\vertex (c1) at (-0.7,-1.5) ;
						\vertex (e) at (3,2.3) {\small{$0, b_1$}};
						\vertex (e1) at (3.,1.5) ;
						\vertex (f1) at (0.7,1.5) ;
						\vertex (h) at (3,-2.3) {\small{$0, b_2$}};
						\vertex (h1) at (3.,-1.5) ;
						\vertex (g1) at (0.7,-1.5) ;
						\diagram*{
							(a1)--[fermion] (b1),
							(d1)--[fermion] (c1),
							(f1)--[fermion] (e1),
							(g1)--[fermion] (h1),
						};
					\end{feynman}
					\filldraw[color=black,thick, fill=gray!15](0,0) ellipse (1.1 and 2.1);
					\begin{feynman}
						\vertex (O) at (0,0){$\widehat{v}_{4,2}$} ;
					\end{feynman}
				\end{tikzpicture} 
			} &\equiv\!\!
			\mathord{
				\begin{tikzpicture}[scale=0.5, baseline=-0.65ex]	
					\begin{feynman}
						\vertex (U) at (-4,2.4) {\small{$x, a_1$}};
						\vertex (V) at (-4,-2.4) {\small{$x, b_1$}};
						\vertex (W) at (4,2.4) {\small{$0, a_2$}};
						\vertex (Z) at (4,-2.4) {\small{$0, b_2$}};
						\vertex (A) at (0,2.5);
						\vertex (B) at (0,-2.5);
						\vertex (z) at (0,2.3);
						\vertex (w) at (0,-2.3);
						\vertex (a1) at (-3.5,1.8) ;
						\vertex (a2) at (-3.5,-1.8) ;
						\vertex (b1) at (-1.1,1.8) ;
						\vertex (d1) at (-3.5,0) ;
						\vertex (c1) at (-1.1,-1.8) ;
						\vertex (e1) at (3.5,1.8) ;
						\vertex (f1) at (1.1,1.8) ;
						\vertex (h1) at (3.5,-1.8) ;
						\vertex (g1) at (1.1,-1.8) ;
						\vertex (j1) at (-1.1, 1.8);
						\vertex (j2) at (-1.1, -1.8);
						\vertex (j3) at (1.1, 1.8);
						\vertex (j4) at (1.1, -1.8);
						\vertex (k1) at (-0.95, 1.65);
						\vertex (k2) at (-0.95, -1.65);
						\vertex (k3) at (0.95, 1.65);
						\vertex (k4) at (0.95, -1.65);
						\diagram*{
							(j1)--[plain,thick](j2),
							(j1)--[plain,thick](j3),
							(j3)--[plain,thick](j4),
							(j4)--[plain,thick] (j2),
							(k1)--[plain,dashed,thick](k2),
							(k1)--[plain,dashed,thick](k3),
							(k3)--[plain,dashed,thick](k4),
							(k4)--[plain,dashed, thick] (k2),
							(a1)--[fermion] (b1),
							(a2)--[fermion] (c1),
							(f1)--[fermion] (e1),
							(g1)--[fermion] (h1),
						};
					\end{feynman}
				\end{tikzpicture} 
			} 
			\!\!\!= g_B^4\,\widehat{C}^{\,a_1 a_2 b_1 b_2}\,\widehat{v}_{4,2}(x,\epsilon)\,\Delta^2(x,\epsilon)
		\end{split} 
		\label{v42B}
	\end{equation}
where the color factor is
	\begin{align}
		\begin{split}
			\widehat{C}^{\,a_1 a_2 b_1 b_2} &=
			\Tr_\cR T^{a_1}T^{b_1}T^{a_2}T^{b_2} -\Tr_{\mathrm{adj}} T^{a_1}T^{b_1}T^{a_2}T^{b_2} ~ 
		\end{split}
		\label{C4prime}
	\end{align}
while the function $\widehat{v}_{4,2}(x,\epsilon)$ is\,\footnote{In \cite{Billo:2019job} this function, originally computed in \cite{Billo:2017glv}, is denoted by $v_{4,2}^{(B)}$ and contains also a factor of $g_B^4$ which here we have put in front of the amplitude.}
	\begin{align}
		\widehat{v}_{4,2}(x,\epsilon)=
		\frac{(\pi x^2)^{2\epsilon}}{(8\pi^2)^2 }\,3\,\zeta(3)+\cO(\epsilon)~.
		\label{v42Bis}
	\end{align}
Inserting this diagram in the two-point correlator, we find 
	\begin{equation}
		\label{v42Bcontribution}
		\begin{split}
			\mathord{
				\begin{tikzpicture}[scale=0.35, baseline=-0.65ex]	
					\newcommand\tmpda{-0.97cm}
					\newcommand\tmpdb{-2.9cm}
					\newcommand\tmpdf{-0.75cm}
					\newcommand\tmpdg{-2.45cm}
					\begin{feynman}
						\vertex (i) at (-4.,0) {$\bullet$};
						\vertex (j) at (4.,0) {$\bullet $ };
						\vertex (a) at (-4.8,0.) {\small{$x$}};
						\vertex (a1) at (-4,0) ;
						\vertex (b1) at (-0.9,0.5) ;
						\vertex (d1) at (0.8,0) ;
						\vertex (c1) at (-1.1,-1.2) ;
						\vertex (e) at (4.8,0.) {\small{$0$}};
						\vertex (e1) at (4,0) ;
						\vertex (f1) at (1.1,1.2) ;
						\vertex (h1) at (4,0) ;
						\vertex (g1) at (1.1,-1.2) ;
						\vertex (o2) at (0,-0.6) {$\vdots$};
						\diagram*{
							(a1) -- [plain, half left, looseness=1.2, with arrow=\tmpda] (h1),
							(a1) -- [plain, half left, looseness=1.2, with arrow=\tmpdb] (h1),
							(a1) -- [plain, half left, looseness=0.65, with arrow=\tmpdf] (h1),
							(a1) -- [plain, half left, looseness=0.65, with arrow=\tmpdg] (h1),
							(a1) -- [fermion, half right, looseness=1.2] (h1),
							(a1) -- [fermion, half right, looseness=0.8] (h1),
						};
					\end{feynman}
					\filldraw[color=black,thick, fill=gray!15](0,2) ellipse (1.1 and 2.1);
					\begin{feynman}
						\vertex (O) at (0,2){$\widehat{v}_{4,2}$} ;
					\end{feynman}
				\end{tikzpicture} 
			} =g_B^{2n+4} \, \,\widehat{v}_{4,2}(x,\epsilon)\,\widehat{\cG}_{\vec{n},\vec{m}}\, \Delta^n(x,\epsilon) 
		\end{split}
	\end{equation}
	with
	\begin{align}
		\widehat{\cG}_{\vec{n},\vec{m}}=n(n-1)\,n!\,
		\widehat{C}^{\,a_1 a_2 b_1 b_2}\,R_{\vec{n}}^{a_1a_2c_1\ldots c_{n-2}}\,
		R_{\vec{m}}^{b_1b_2c_1\ldots c_{n-2}} ~.
		\label{Gprime}
	\end{align}
The symmetry prefactor is due to the fact that there are two ways to connect two of the $n$ indices of $R_{\vec{n}}$ with $\widehat{C}$, two ways to connect two of the $n$ indices of $R_{\vec{m}}$ with $\widehat{C}$, and $(n-2)!$ ways to contract the remaining $(n-2)$ indices of $R_{\vec{n}}$ and $R_{\vec{m}}$ among themselves. Therefore, altogether we have
\begin{align}
2\,\binom{n}{2}\times 2\,\binom{n}{2}\times (n-2)!=n(n-1)\,n!~.
\end{align}
We point out that in contrast to the previous cases, the color factor (\ref{Gprime}) is not proportional to the tree-level coefficient $G^{(0)}_{\vec{n},\vec{m}}$ and must computed on a case-by-case basis. 

Combining all the two-loop contributions together, we obtain
	\begin{align}
		\label{G2loop}
		\begin{split}
			\cG_{\vec{n},\vec{m}}^{(2)}(x,\epsilon)&=\frac{n(n+1)}{2}\,v^2_{2,1}(x,\epsilon)+n\big[v_{2,2}(x,\epsilon)-v_{4,2}(x,\epsilon)\big]\\
			&\qquad-n(n-2)\,v_{2,1}(x,\epsilon)\,e_{2,1}(x,\epsilon)
			+\widehat{v}_{4,2}(x,\epsilon)\,\frac{\widehat{\cG}_{\vec{n},\vec{m}}}{G_{\vec{n},\vec{m}}^{(0)}}+ \cO(\epsilon)~.
		\end{split}
	\end{align}
These terms define the difference-theory part $\cG^{\prime\,(2)}$ while the $\cO(\epsilon)$ term is the 
two-loop evanescent contribution $\delta^{\prime}\cG^{(2)}$. Since this is relevant only at the next perturbative order, its explicit expression is not needed. Note that the combination $[v_{2,2}(x,\epsilon)- v_{4,2}(x,\epsilon)]$ is finite when $\epsilon\to 0$, as one can see from (\ref{v22is}) and \eqref{v42Ais}. This means that the UV singularities are entirely encoded in the one-loop function $v_{2,1}(x,\epsilon)$, as expected from the one-loop exactness of the $\beta$-function.

\subsection{Renormalized correlators}
\label{subsecn:renorm}
Collecting all contributions up to two loops, we find that the correlator (\ref{OnOm}) is
\begin{align}
		\label{Gup to2l}
			G_{\vec{n},\vec{m}}(g_b,x,\epsilon)&
			=G_{\vec{n},\vec{m}}^{(0)}\,\bigg[\frac{g_B^{2n}}{\big(1-g_B^2\,v_{2,1}(x,\epsilon)\big)^n} - g_B^{2n+2}\, n\, e_{2,1}(x,\epsilon)\notag\\[2mm]
			&\quad
		+g_B^{2n+4}\,n\big[v_{2,2}(x,\epsilon)-v_{4,2}(x,\epsilon)\big]
		-g_B^{2n+4}\,n\,(n-2)\,v_{2,1}(x,\epsilon)\,e_{2,1}(x,\epsilon)\notag\\[2mm]
		&\quad
			+g_B^{2n+4} \, \widehat{v}_{4,2}(x,\epsilon)\,
			\frac{\widehat{\cG}_{\vec{n},\vec{m}}}{G_{\vec{n},\vec{m}}^{(0)}}\bigg]+\cO(g_B^{2n+6})~.
	\end{align} 
In this expression, the terms proportional to $g_B^{2n+4}$ are finite in the limit $\epsilon\to0$, while the UV singularities are captured by the geometric progression
\begin{equation}
	\label{eq:divergences}
	\frac{1}{\big(1-g_B^2\,v_{2,1}(x,\epsilon)\big)^n}= 1 + n\, g_B^2 \,v_{2,1}(x,\epsilon) +\frac{1}{2}n(n+1) 
	\,g_B^4\, v_{2,1}(x,\epsilon)^2+ \ldots~.
\end{equation} 
To remove these divergences, we use a modified minimal subtraction scheme and introduce the renormalized coupling constant $g_*$ evaluated at the scale $1/\mu$ according to\,\footnote{In (\ref{eq:renormalized coupling}) we have included the factor $\mathrm{e}^{-\frac{\epsilon}{2}(2+\gamma+\log\pi)}$ in order to remove some numerical terms in the following formulas.}
\begin{align}
g_B = g_* \big(\mu\,\mathrm{e}^{-\frac{1}{2}(2+\gamma+\log\pi)}\big)^{\epsilon} Z_{g_*} \,\equiv \,
 g_* \,\bar{\mu}^{\epsilon}\,Z_{g_*} 
\label{eq:renormalized coupling}
\end{align}
where $\gamma$ is the Euler-Mascheroni constant and
\begin{equation}
	\begin{split}
Z^2_{g_*}  =\bigg(1+\frac{g_*^2\,\beta_0}{16\pi^2\epsilon}\bigg)^{-1}~.
\label{Zg}
	\end{split}
\end{equation}
Then, the renormalized correlator $G^*_{\vec{n},\vec{m}}$ is obtained by replacing in (\ref{Gup to2l}) the bare coupling with the renormalized one according to (\ref{eq:renormalized coupling}) and taking the limit $\epsilon\to 0$, namely\,\footnote{For composite operators a separate renormalization function is in general required. The reason why in our case we can reabsorb all singularities with a single function $Z_{g_*}$ is because the operators $O_{\vec{n}}$ have an overall factor of $g_B^n$ as one can see from (\ref{On}).}
\begin{equation}
	\label{G*}
	G^*_{\vec{n},\vec{m}} =\lim_{\epsilon\to 0} G_{\vec{n},\vec{m}}(g_B,x,\epsilon)\Big|_{g_B = g_* \,\bar{\mu}^{\epsilon}\,Z_{g_*} }~.
\end{equation} 
It is easy to check that in this way the $\epsilon$-poles of $Z_{g_*}$ precisely remove those appearing in (\ref{eq:divergences}), leading to a finite result in $d=4$. Indeed, one has
\begin{equation}
	\begin{split}
		\lim_{\epsilon\to 0}\frac{g_B^{2n}}{\big(1-g_B^2\,v_{2,1}(x,\epsilon)\big)^n} \bigg|_{g_B = g_* \,\bar{\mu}^{\epsilon}\,Z_{g_*} } &~=\bigg( \frac{g_*^{2}}{1-g_*^2\frac{\beta_0}{16\pi^2}\log({\mu}^2 x^2) }\bigg)^n
		\,\equiv\,g^{2n}
	\end{split}
	\label{gx}
\end{equation} 
where $g$ is the running coupling  evaluated at the scale $|x|$. Furthermore, using (\ref{eq:e21is}) one can show that
\begin{equation}
\lim_{\epsilon\to 0} g_B^{2n+2}\, n\, e_{2,1}(x,\epsilon)\Big|_{g_B=g_* \,\bar{\mu}^{\epsilon}\,Z_{g_*} } =-g^{2n+4}\,\frac{3N\zeta(3)\beta_0}{64\pi^4}\,n(n+1)+\cO\big(g^{2n+6}\big)~. 
\label{evan}
\end{equation}
Here we explicitly see that the one-loop evanescent contribution $e_{2,1}$, upon renormalization, produces a finite term at two loops.
As already anticipated, the remaining contributions of order $g_B^{2n+4}$ in (\ref{Gup to2l}) are regular in the limit $\epsilon\to 0$. So for them we can safely set $\epsilon=0$ and replace the bare coupling with the running coupling $g$. In particular, we see that 
\begin{align}
\lim_{\epsilon\to 0} g_B^{2n+4}\, n(n-2)\,v_{2,1}(x,\epsilon)\, e_{2,1}(x,\epsilon)\Big|_{g_B=g_* \,\bar{\mu}^{\epsilon}\,Z_{g_*} } &\!\!=g^{2n+4}\,\frac{3N\zeta(3)\beta_0}{64\pi^4}\,n(n-2)
+\cO\big(g^{2n+6}\big)~.
\label{evan1}
\end{align}
The two contributions (\ref{evan}) and (\ref{evan1}) produced by the evanescent function $e_{2,1}$ are new with respect to the analysis of \cite{Billo:2019job}.

Altogether, the renormalized correlator (\ref{G*}) takes the following simple form
\begin{equation}
	\begin{split}
		G^*_{\vec{n},\vec{m}} &=g^{2n} \,G_{\vec{n},\vec{m}}^{(0)} \,\bigg[ 1 +\Big(\frac{g^2}{8\pi^2}\Big)^{\!2}\, 3\zeta(3)\,\cC^{(2)}_{\vec{n},\vec{m}}+\cO(g^6)\bigg]
		\label{Gnmren}
	\end{split} 
\end{equation}
where
\begin{equation}
	\begin{split}
		\cC^{(2)}_{\vec{n},\vec{m}}=2n\,\big(C_\cR\,i_\cR-N^2\big) +\frac{\widehat{\cG}_{\vec{n},\vec{m}}}{G_{\vec{n},\vec{m}}^{(0)}}
		\label{Gnmren1}
	\end{split} 
\end{equation}
with $\widehat{\cG}_{\vec{n},\vec{m}}$ defined in (\ref{Gprime}). 
We would like to remark that in (\ref{Gnmren}) there is a hidden dependence on $x$ since the running coupling $g$ is evaluated at the scale $|x|$. Of course, in conformal theories, where $\beta_0=0$ and the coupling does not run, this dependence is absent.

For completeness and to facilitate the comparison with existing results in the literature, we now give the explicit expressions of the renormalized correlator (\ref{Gnmren}) for the lowest dimensional operators when the representation $\cR$ is  (\ref{eq:Ris}). Evaluating $\widehat{\cG}_{\vec{n},\vec{m}}$ in this case and using the results in (\ref{G0;23}), we obtain
\begin{subequations}
\begin{align}
G^*_{2,2} &=g^{4} \,\frac{N^2-1}{2}\,\bigg\{ 1 +\Big(\frac{g^2}{8\pi^2}\Big)^{\!2}\,\, \frac{3\zeta(3)}{2N}\,\big[N_f (2 N^2-3) -10 N^3\label{G22ren}\\
&\qquad+N_a(N-2)(5 N^2-6 N-12) +N_s(N+2)
   (5 N^2+6 N-12) \big]+\cO(g^6)\bigg\}~,\notag\\[2mm]
  G^*_{3,3} &=g^{6}\, \frac{3(N^2-1)(N^2-4)}{8N}\,\bigg\{ 1 +\Big(\frac{g^2}{8\pi^2}\Big)^{\!2}
  \,\, \frac{9\zeta(3)}{2N}\,\big[N_f (N^2-3)-4 N^3\label{G33ren}\\
&\qquad+2 N_a(N^3-4 N^2+12)+2 N_s (N^3+4 N^2-12)\big]+\cO(g^6)\bigg\}~.\notag
\end{align}
\label{G2233ren}%
\end{subequations}
The expressions for the correlators of operators with dimensions 4 and 5 
can be found in Appendix~\ref{App:Gmn}.

\section{Matrix Model approach}
\label{sec:MM}
In this section, we turn our attention to the calculation of the chiral/anti-chiral correlator (\ref{OnOm}) by supersymmetric localization.
	
In four-dimensional gauge theories with extended supersymmetry algebra,  this technique was originally applied in  \cite{Pestun:2007rz} to compute the   partition functions and the vacuum expectation value of half-BPS  Wilson loops on $S^4$ in terms of an interacting matrix model. Subsequently, supersymmetric localization on the four-sphere was also applied to extremal correlators of chiral/anti-chiral operators in superconformal $\cN=2$ theories \cite{Baggio:2014ioa,Baggio:2014sna,Gerchkovitz:2016gxx}. 
For all these observables,  the path-integral  localizes about a set of critical configurations,  known as the BPS {locus}, in which the vector-multiplet scalar $\phi$ reduces to a constant Hermitian traceless matrix $a$, while the gauge field  reduces to an instanton solution. As a result,  the ``sum'' over the critical points corresponds to both an integral over the matrix $a$ and a sum over instantons. Moreover, the saddle-point approximation, which takes into account the fluctuation determinants around the BPS locus, is exact. 

However, to avoid infrared divergences that obstruct this computation, the  theory must be placed on a compact space,  such as a four sphere $S^4$ of radius $R$, or a squashed version of it. For this reason, one might expect that localization techniques can provide information on observables in flat space only for conformal theories, since  a mapping from $S^4$ to $\mathbb{R}^4$ is needed. However, contrarily to these expectations and common lore, we demonstrate below that, within a specific regime of validity, localization on the sphere still retains a meaningful connection with flat space when the conformal symmetry is broken at the quantum level.

As explained in \cite{Pestun:2007rz, Billo:2024fst,Billo:2023igr}, the matrix model of a theory with non-vanishing $\beta$-function requires a particular regularization prescription for the fluctuation determinants. Usually, these appear  as infinite products that have to be regulated. In particular, if the gauge theory has a zero $\beta$-function, {\it{i.e.}} $i_\cR = N$, these products are packaged into well-behaved $H$-function (see (\ref{eq:H is})); if the matter hypermultiplets are massive, a divergent factor remains, but it is $a$-independent and consequently, it disappears in properly normalized observables. 

However, when the theory is asymptotically free, i.e. $i_\cR < N$, the regularization based on the $H$-functions is not sufficient. In this case, the divergences are removed by  ``embedding'' 
 the theory in a larger  model  with matter in a representation $\cR^*$ such that $\cR\subset \cR^*$ and $i_{\cR^*} = N$, so that the divergent terms cancel. Let $g_*$ be the coupling constant of the regulated theory and let the additional hypermultiplets (those not in $\cR$) have mass $M$. In the decoupling limit $M R\to \infty$, one remains with the original  theory based on $\cR$, but the additional matter affects the coefficient of the Gaussian term in the matrix integral,  which becomes proportional to the combination
\begin{equation}
	\label{runnmm}
		\frac{1}{g_*^2} - \frac{\beta_0}{16\pi^2} \log ( M^2 R^2)
\end{equation}
where $\beta_0=2(N-i_\cR)$.

\begin{figure}[h]
\begin{center}
\includegraphics[scale=0.28]{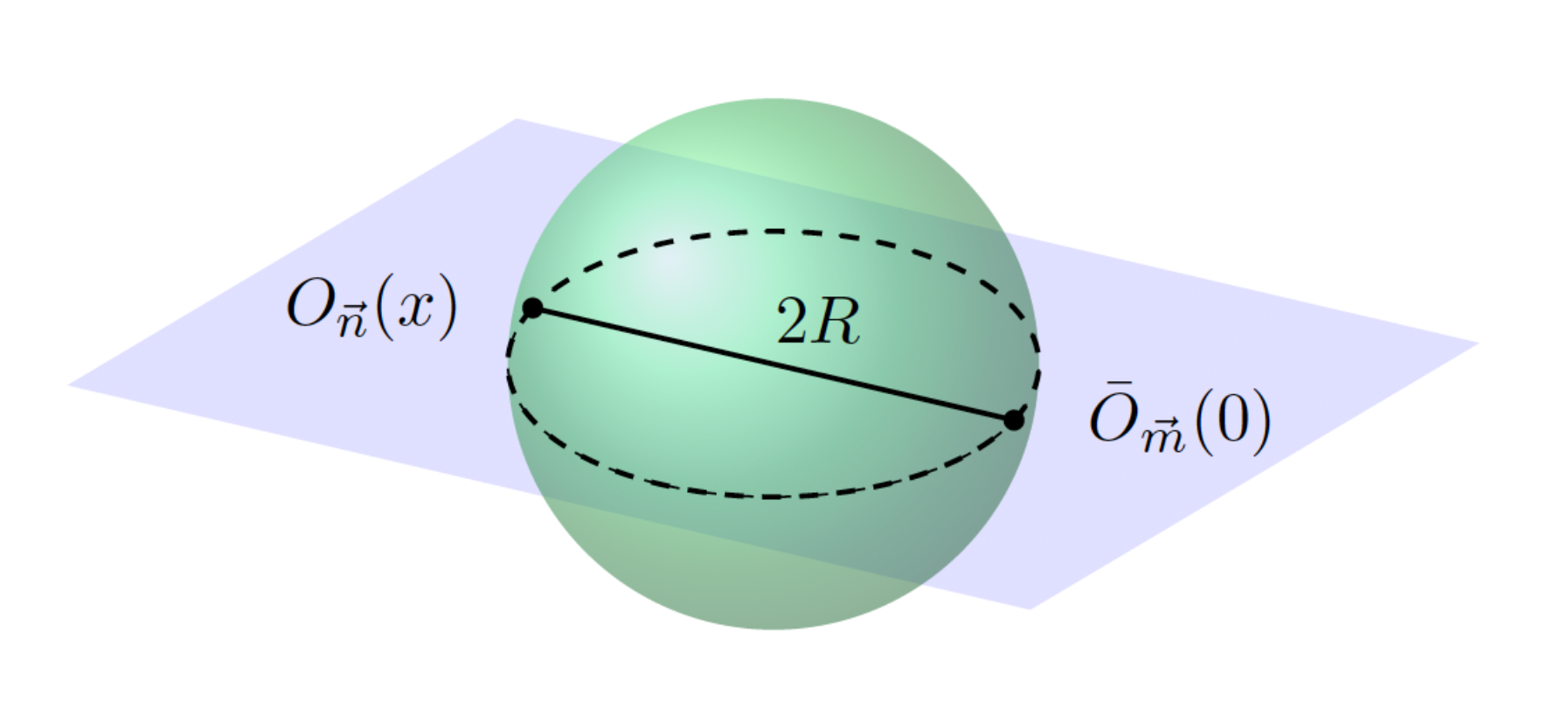}%
\end{center}
\vspace{-0.5cm}
\caption{A pictorial representation of the relation between $S^4$ and $\mathbb{R}^4$ which shows that operators at antipodal points on $S^4$ correspond to operators at a separation $|x|=2R$ in $\mathbb{R}^4$.}
	\label{FD}
\end{figure}

The observables we want to compute with localization must
correspond to BPS configurations on $S^4$.
For instance, the $\frac{1}{2}$-BPS Wilson loop  must lie on a great circle, while for the two-point functions we require  that the chiral and anti-chiral operators be placed at antipodal points, as represented in Fig.\,\ref{FD}. 
Therefore, the two-point function on $S^4$ maps to a field-theory correlator at a separation $|x| = 2R$ on $\mathbb{R}^4$. This relation sets the scale of this observable in the two descriptions. Furthermore, if we choose $M=2\mu$, we see that the combination (\ref{runnmm}) precisely coincides with the running coupling
constant $g$ at the scale $|x|$ introduced in Section\,\ref{subsecn:renorm}, namely
\begin{equation}
	\label{runft}
	\frac{1}{g_*^2} - \frac{\beta_0}{16\pi^2} \log (\mu^2 x^2) = \frac{1}{g^2}~,  
\end{equation}
in full agreement\,\footnote{Note that in the decoupling limit the factor of 2 between the UV scales $M$ and $\mu$ is immaterial since $\log (M^2 x^2) = \log (\mu^2 x^2) + \log 4$ and the shift by $\log 4$ is negligible with respect to the large logs.} with (\ref{gx}).

Another effect of adding additional massive matter hypermultiplets is that an instanton configuration with instanton number $k$ is weighted with $\Lambda^{k\beta_0}$ where 
\begin{equation}
	\Lambda = \mu\, \ee^{-\frac{8\pi^2}{g_*^2 \beta_0}}
	\label{Lambdadef}
\end{equation}  
is the mass scale at which the running coupling diverges. 

Taking all this into account, the partition of the matrix model arising from localization is 
\begin{equation}
	\label{partfnc}
		\cZ = \int \!Da\,\, \ee^{- \frac{8\pi^2}{g^2}\,R^2\tr a^2} \,\big|Z_{\mathrm{1-loop}}(a R)\big|^2 \,
		\big|Z_{\text{inst}}(aR, \Lambda^{\beta_0})\big|^2~
\end{equation}
where $Z_{\mathrm{1-loop}}$ represents the determinant of the fluctuations around the fixed points and $Z_{\text{inst}}$ is the Nekrasov instanton partition function encoding the result of the integration over the instanton moduli. In the regime 
\begin{equation}
	\label{regime}
		M \gg 1/R \gg \Lambda~,
\end{equation}
the additional matter hypermultiplets decouple and instantons can be neglected\,\footnote{The Nekrasov  instanton partition function can be worked out also for the $\mathcal{N}=2$ theories we are considering. However, as clear from (\ref{Lambdadef}), it yields non-perturbative contributions which, in the regime (\ref{regime}), are exponentially suppressed with respect to the perturbative terms we compute. We stress these non-perturbative corrections depend on the dimensionless product $\Lambda R$ and have an intrinsic \emph{infrared} nature. As a result, we expect that such contributions differ between the sphere and flat space.}, so that the partition function (\ref{partfnc}) reduces to
\begin{equation}
	\label{partfnc2}
	\cZ = \int \!Da\,\, \ee^{- \frac{8\pi^2}{g^2} R^2\tr a^2 - S_{\mathrm{int}}(a R)}~ 
\end{equation}
where
\begin{equation}
		S_{\mathrm{int}}(a R) = - \log \big|Z_{\mathrm{1-loop}}(a R)\big|^2~. 
\end{equation}
From the results of \cite{Pestun:2007rz}, one can show that this interaction action can be written as
\begin{align}
	\label{Sintis}
	S_{\mathrm{int}}(a R) = \Tr^\prime \log H(a R)
\end{align}
where we have introduced the ``difference theory'' trace \cite{Billo:2019fbi}
\begin{align}
\Tr^\prime \bullet \,\equiv \, \Tr_\cR \bullet \,-\,\Tr_{\mathrm{adj}}\bullet
\end{align}
and defined 
\begin{equation}
	\label{eq:H is}
	H(x)= G(1+\ii x) \,G(1-\ii x)\, \ee^{-(1+\gamma)x^2}
	=  \prod_{n=1}^{\infty}\left(1+\frac{x^2}{n^2}\right)^n \ee^{-\frac{x^2}{n}}~,
\end{equation} 	
with $G$ being the Barnes $G$-function. Expressing the matrix $a$ as  
\begin{equation}
	\label{rescalea}
	a =\sqrt{\frac{g^2}{8\pi^2}} \,\frac{\hat a}{R}
\end{equation}
in such a way that $\hat a$ is dimensionless, the Gaussian term in (\ref{partfnc2}) becomes
$\exp(-\tr \hat a^2)$ and the coupling constant appears in the interaction vertices of $S_{\mathrm{int}}(\hat a)$. 
Furthermore, the partition function gets rescaled by an overall factor which, however, is irrelevant in any normalized correlator and consequently, it can be ignored. Writing for simplicity again $a$ instead of $\hat a$, we finally have
\begin{equation}
	\label{partfnc3}
		\cZ = \int\! Da\,\, \ee^{- \tr a^2 - S_{\mathrm{int}}(a)}~. 
\end{equation}
Exploiting the expansion of (the logarithm of) the Barnes $G$-functions for small arguments, one can easily deduce the following form of the interaction action
\begin{equation}
	\label{Seintis2}
	S_{\mathrm{int}}(a) =- \sum_{m=2}^\infty (-1)^m\Big(\frac{g^2}{8\pi^2}\Big)^{\!m} \,\frac{\zeta(2m-1)}{m}\,\Tr^\prime a^{2m}
	= -\Big(\frac{g^2}{8\pi^2}\Big)^{\!2}\, \frac{\zeta(3)}{2}\, \Tr^\prime a^4 + \cO(g^6)~.
\end{equation}
We remark that all explicit dependence on $R$ has disappeared. However, there remains a hidden dependence on $R$ through the running coupling $g$ which, as argued above, is evaluated at the scale $|x|=2R$.

In the following, we adopt the so-called ``full Lie algebra'' approach, originally introduced in \cite{Billo:2017glv}, namely we expand the matrix $a$ over a basis of generators of $\mathfrak{su}(N)$ in the fundamental representation according to $a=a^b\,T^b$ ($b=1,\ldots,N^2-1$), and integrate over all components $a^b$ with a flat integration measure
\begin{equation}
	\label{Daflat}
		\int \!Da = \int \prod_{b=1}^{N^2-1} \frac{da^b}{\sqrt{2\pi}}~.
\end{equation}
The normalization is chosen so that the partition function and the basic ``contraction'' in the free Gaussian model, labeled by a subscript $0$, are simply
\begin{equation}
	\label{contraa}
		\cZ_0 = \int\!Da\, \, \ee^{-\tr a^2} = 1 \qquad\mbox{and}\qquad
		\big\langle a^b\, a^c\big\rangle_0 \equiv \int\!Da\,\,a^b\, a^c\, \ee^{-\tr a^2} = \delta^{bc}~.
\end{equation}

Given an arbitrary operator $f(a)$, its vacuum expectation value in the interacting matrix model representing the $\cN=2$ gauge theory is
\begin{equation}
	\label{vevf}
		\big\langle f(a) \big\rangle = \frac{1}{\cZ} \int\!Da\,\, f(a)\, \ee^{-\tr a^2 - S_{\mathrm{int}}(a)} 
		= \frac{\big\langle f(a)\,\ee^{- S_{\mathrm{int}}(a)}\big\rangle_0\phantom{\Big|}}{\big\langle 
		\ee^{- S_{\mathrm{int}}(a)} \big\rangle_0\phantom{\Big|}}~.
\end{equation}
In perturbation theory, it can be computed diagrammatically: expanding the interaction action $S_{\mathrm{int}}(a)$ in powers of $g$ according to (\ref{Seintis2}) provides the vertices
while the propagators are simply given by the free contraction (\ref{contraa}). For instance, at order $g^4$ there is a quartic interaction vertex proportional to $\zeta(3)$:
\begin{equation}
	\label{quarticv}
	\Tr^\prime a^4
\qquad \longleftrightarrow \qquad  \mathord{
		\begin{tikzpicture}[radius=2.cm, baseline=-0.65ex,scale=0.5]
			\begin{feynman}
				\vertex (A) at (1,0);
				\vertex (B) at (-1,0);
				\vertex (C) at (0,1);
				\vertex (D) at (0,-1);
				\vertex (O) at (0,0);
				\diagram*{
					(A) --[ plain,thick] (O),
					(D) --[ plain,thick] (O),
					(B)--[plain,thick] (O),
					(C)--[plain,thick] (O),
				};
			\end{feynman}
			\node at (0,0)[rectangle,fill,inner sep=3.1pt]{};
	\end{tikzpicture} }  ~.
\end{equation} 
This treatment very closely resembles the computation of a corresponding observable in field theory by means of Feynman diagrams and is therefore well suited to check matrix-model and field-theory results against each other.

\subsection{Chiral/anti-chiral correlators at order \texorpdfstring{$g^4$}{}}
\label{subsec:cac4}

In \cite{Baggio:2014ioa,Baggio:2014sna,Gerchkovitz:2016gxx,Billo:2017glv} it was shown that for conformal theories the correlators of chiral/anti-chiral operators in $\mathbb{R}^4$ are captured by analogue correlators in the matrix model on $S^4$. Here we show that, up to two loops, the same is true in models where conformal symmetry is broken by a non-vanishing $\beta$-function. 

In the matrix model, both the chiral operators $O_{\vec n}(x)$ introduced in (\ref{On}) and their anti-chiral counterparts are represented by 
\begin{equation}
	\label{defcO}
		\cO_{\vec n}(a) =\,\,\, :\prod_{i=1}^\ell g^{n_i} \tr a^{n_i}:\,\, = g^n\,R_{\vec{n}}^{a_1\ldots a_n}\,:a^{a_1}\ldots a^{a_n}: 
\end{equation}
where the fully symmetric $R$ tensor is defined as in (\ref{Rtensor}) and the normal ordering ensures that $\cO_{\vec n}(a)$ has no self-interactions, just as $O_{\vec n}(x)$. It is easy to realize that this way of representing the operators in the matrix model is completely equivalent to the most commonly used one in which the mixing with lower dimensional operators is taken into account and the coefficients are fixed with the Gram-Schmidt orthonormalization procedure to eliminate all self-contractions.

Then, according to (\ref{vevf}), the two-point function correlator is given by
\begin{align}
G_{\vec{n},\vec{m}}= \frac{\big\langle \cO_{\vec n}(a)\,\cO_{\vec m}(a)\,\ee^{- S_{\mathrm{int}}(a)}\big\rangle_0\phantom{\Big|}}{\big\langle 
		\ee^{- S_{\mathrm{int}}(a)} \big\rangle_0\phantom{\Big|}}
	\label{GnmMM}
\end{align}
which up to order $g^4$ explicitly reads
\begin{align}
G_{\vec{n},\vec{m}}=\big\langle \cO_{\vec n}(a)\,\cO_{\vec m}(a)\big\rangle_0+\Big(\frac{g^2}{8\pi^2}\Big)^{\!2}
\,\frac{\zeta(3)}{2}\,\big\langle \cO_{\vec n}(a)\,\cO_{\vec m}(a)\,\Tr^\prime a^4\big\rangle_0^c\,+\,\cO(g^6)~.
\label{GnmMMexp}
\end{align}
Here we have used (\ref{Seintis2}) and denoted by a superscript $c$ the connected part of the vacuum expectation value, namely
\begin{align}
\big\langle \cO_{\vec n}(a)\,\cO_{\vec m}(a)\,\Tr^\prime a^4\big\rangle_0^c\,\equiv\,
\big\langle \cO_{\vec n}(a)\,\cO_{\vec m}(a)\,\Tr^\prime a^4\big\rangle_0-\big\langle \cO_{\vec n}(a)\,\cO_{\vec m}(a)\big\rangle_0\,\big\langle\Tr^\prime a^4\big\rangle_0~.
\label{connected}
\end{align}
The first contribution in the right-hand side of (\ref{GnmMMexp}) can be represented by the following diagram
\begin{equation}
	\label{eq:tree-level mm}
	\mathord{
		\begin{tikzpicture}[scale=0.7, baseline=-0.65ex]
			\begin{feynman}
				\vertex (A) at (-2,0) ;
				\vertex (a) at (-2,0) {$\bullet$};
				\vertex (c) at (2,0) {$\bullet $ };
				\vertex (C) at (2,0) ;
				\vertex (V) at (3.1,0) {\small{$\cO_{\vec{m}}(a)$} };
				\vertex (V) at (-3,0) {\small{$\cO_{\vec{n}}(a)$} };
				\vertex (o) at (0,0) {$.$};
				\vertex (o1) at (0,0.3) {$.$};
				\vertex (o2) at (0,-0.3) {$.$};
				\diagram*{
					(A) -- [plain, half left, looseness=0.85] (C),
					(A) --[plain, half right, looseness=0.85] (C),
					(A) --[plain, half right, looseness=0.55] (C),
					(A) --[plain, half left, looseness=0.55] (C),
				};
			\end{feynman}
		\end{tikzpicture} ~.
	} 
\end{equation}
Inserting the explicit expressions (\ref{defcO}) of the operators, it can be trivially evaluated using Wick's theorem with the contractions (\ref{contraa}) and the result is
\begin{align}
\big\langle \cO_{\vec n}(a)\,\cO_{\vec m}(a)\big\rangle_0=g^{2n}\,n!\,R_{\vec n}^{a_1\ldots a_n}\, R_{\vec m}^{a_1 \ldots a_n} =g^{2n}\,G^{(0)}_{\vec{n},\vec{m}}
\label{treeMM}
\end{align}
where $G^{(0)}_{\vec{n},\vec{m}}$ is the tree-level term of the field-theory computation (see (\ref{eq:cal G0 is})).
 
Let us now consider the interacting part. There are two important points that need to be taken into account. The first is that the quartic vertex (\ref{quarticv}) can be written as
\begin{equation}
\Tr^\prime a^4
=\widehat{C}_s^{abcd}\,a^{a}a^{b}a^{c}a^{d} 
\label{quarticv2}
\end{equation}
where $\widehat{C}_s^{abcd} = \Tr^\prime\left(T^{(a} T^b T^c T^{d)}\right)$ is the \emph{completely symmetrized} version of the four-index tensor defined in (\ref{C4prime}), namely\,
\footnote{Since $\widehat{C}^{\ell_1\ell_2\ell_3\ell_4}$ is a trace of four generators it is cyclically symmetric; hence the six permutations exhibited in (\ref{Chatsymm}) are enough to fully symmetrize in all indices.}
\begin{align}
\widehat{C}_s^{abcd}=\frac{1}{6}\Big(\widehat{C}^{abcd}+
\widehat{C}^{acdb}+\widehat{C}^{adbc}+\widehat{C}^{abdc}+\widehat{C}^{acbd}+\widehat{C}^{adcb}
\Big)~.
\label{Chatsymm}
\end{align}
The second point is that, in contrast to the operators $\cO_{\vec n}(a)$ and $\cO_{\vec m}(a)$, the quartic interaction is \emph{not} normal ordered and consequently,  contractions between its legs are possible. Taking these observations into account, it is not difficult to realize that the connected contribution (\ref{connected}) corresponds to the sum of two diagrams.  
The first diagram is given by
\begin{align}
\mathord{
		\begin{tikzpicture}[scale=0.7, baseline=-0.65ex]
			\begin{feynman}
			\vertex (a) at (-2,0) {$\bullet$};
				\vertex (c) at (2,0) {$\bullet $ };
				\vertex (A) at (-2,0) ;
				\vertex (C) at (2,0) ;
				\vertex (o) at (0,0) {$.$};
				\vertex (o1) at (0,0.3) {$.$};
				\vertex (o2) at (0,-0.3) {$.$};
				\vertex (f) at (0,0.95) ;
				\vertex (f2) at (0,2.2)  ;
			\vertex (V) at (3.1,0) {\small{$\cO_{\vec{m}}(a)$} };
				\vertex (V) at (-3,0) {\small{$\cO_{\vec{n}}(a)$} };
				\diagram*{
					(A) -- [plain, half left, looseness=0.85] (C),
					(A) --[plain, half right, looseness=0.85] (C),
					(A) --[plain, half right, looseness=0.55] (C),
					(A) --[plain, half left, looseness=0.55] (C),
					(f2) -- [plain,half left] (f),
					(f2) -- [plain,half right] (f),
				};
			\end{feynman}
			\node at (0,0.96)[rectangle,fill,inner sep=3.1pt]{};
		\end{tikzpicture} 
	} = g^{2n}\,12n\,n!\,\widehat{C}_s^{a_1b_1 d \,d}\,\,R_{\vec n}^{a_1c_1\ldots c_{n-1}}\, R_{\vec m}^{b_1c_1 \ldots c_{n-1}}~,
		\label{MM:diagramA}
\end{align}
where we performed all contractions and computed the symmetry factor as follows. There are $2\,\binom{4}{2}$ ways of selecting an ordered pair of indices $(a_1b_1)$ in the tensor $\widehat{C}_s$, $n$ ways of connecting one index $a_1$ of $R_{\vec n}$ with $\widehat{C}_s$, $n$ ways of connecting one index $b_1$ of $R_{\vec m}$ with $\widehat{C}_s$ and
$(n-1)!$ ways of contracting the remaining indices of $R_{\vec n}$ and $R_{\vec m}$ with each other.
This counting gives
\begin{align}
2\,\binom{4}{2}\times n \times n\times (n-1)! = 12 n\,n!
\end{align}
as written in (\ref{MM:diagramA}).  The second diagram is
\begin{align}
	\label{MM:diagramB}
\mathord{
			\begin{tikzpicture}[scale=0.7, baseline=-0.65ex]
				\newcommand\tmpda{-0.45cm}
				\newcommand\tmpdb{-2.85cm}
				\begin{feynman}
					\vertex (A) at (-2,0) ;
					\vertex (C) at (2,0) ;
					\vertex (F) at (0,0) ;
					\vertex (V) at (3.1,0) {\small{$\cO_{\vec{m}}(a)$} };
				\vertex (V) at (-3,0) {\small{$\cO_{\vec{n}}(a)$} };
						\vertex (f1) at (0,-0.5) {$.$ } ;
						\vertex (f2) at (0,-0.7) {$. $ } ;
						\vertex (f3) at (0,-0.9) {$.$ } ;
					\vertex (H) at (0,-1);
					\vertex (a) at (-2,0) {$\bullet$};
				\vertex (c) at (2,0) {$\bullet $ };
					\diagram*{
						(A) -- [plain, half left, looseness=0.9] (F),
						(A) -- [plain, half right, looseness=0.9] (F),
						(F) -- [plain, half left,looseness=0.9] (C),
						(F) -- [plain, half right, looseness=0.9] (C),
						(A) -- [plain, half right, looseness=0.9] (C),
						(A) -- [plain, half right, looseness=1.1] (C),
						(A) -- [plain, half right, looseness=1.4] (C),
					};
				\end{feynman}
				\node at (0,0)[rectangle,fill,inner sep=3.1pt]{};
			\end{tikzpicture} 
		}\\[-1.2cm]
		&=g^{2n}\,6n(n-1)\,n!\,\widehat{C}_s^{a_1a_2b_1b_2}\,\,R_{\vec n}^{a_1a_2c_1\ldots c_{n-2}}\, R_{\vec m}^{b_1b_2c_1 \ldots c_{n-2}}~.\notag\\[-0.2cm]
		\notag
\end{align}
In this case the symmetry factor arises because there are $\binom{4}{2}$ ways of dividing the four indices of $\widehat{C}_s$ into two pairs, two ways of contracting two of the $n$ indices of $R_{\vec n}$ with
$\widehat{C}_s$, two ways of contracting two of the $n$ indices of $R_{\vec m}$ with
$\widehat{C}_s$ and $(n-2)!$ ways of contracting the remaining indices of $R_{\vec n}$ and $R_{\vec m}$ with each other. Altogether, we have
\begin{align}
\binom{4}{2}\times 2\,\binom{n}{2}\times 2\,\binom{n}{2}\times (n-2)! = 6n(n-1)\,n!
\end{align}
as written in (\ref{MM:diagramB}).

Adding the two contributions (\ref{MM:diagramA}) and (\ref{MM:diagramB}), we get
\begin{align}
\big\langle \cO_{\vec n}(a)\,\cO_{\vec m}(a)\,\Tr^\prime a^4\big\rangle_0^c&=6 g^{2n}\Big[
2n\,n!\,\widehat{C}_s^{a_1b_1 d \,d}\,\,R_{\vec n}^{a_1c_1\ldots c_{n-1}}\, R_{\vec m}^{b_1c_1 \ldots c_{n-1}}\notag\\
&\qquad+
n(n-1)\,n!\,\widehat{C}_s^{a_1a_2b_1b_2}\,\,R_{\vec n}^{a_1a_2c_1\ldots c_{n-2}}\, R_{\vec m}^{b_1b_2c_1 \ldots c_{n-2}}
\Big]~.
\label{result1MM}
\end{align}
In Appendix\,\ref{App:MM}, we prove the following relations:
\begin{subequations}
\begin{align}
2n\,n!\,\widehat{C}_s^{a_1b_1 d \,d}\,\,&R_{\vec n}^{a_1c_1\ldots c_{n-1}}\, R_{\vec m}^{b_1c_1 \ldots c_{n-1}}
=
2n\,\Big(C_\cR\,i_\cR-N^2+\frac{N \beta_0}{12}\Big)\,G^{(0)}_{\vec{n},\vec{m}}~,\label{relationA}\\[2mm]
n(n-1)\,n!\,\widehat{C}_s^{a_1a_2b_1b_2}\,\,&R_{\vec n}^{a_1a_2c_1\ldots c_{n-2}}\,R_{\vec m}^{b_1b_2c_1 \ldots c_{n-2}}=\widehat{\cG}_{\vec{n},\vec{m}}-n\,\frac{N \beta_0}{6}\,G^{(0)}_{\vec{n},\vec{m}}~
\label{relationB}
\end{align}
\label{relationsMM}%
\end{subequations}
where $\widehat{\cG}_{\vec{n},\vec{m}}$ is defined in (\ref{Gprime}) and $G^{(0)}_{\vec{n},\vec{m}}$ is the tree-level coefficient. Notice that when we sum these two contributions as in the square brackets of (\ref{result1MM}),
the terms proportional to $N \beta_0$ cancel against each other.

Using the relations (\ref{relationsMM}), reinstating the prefactor 
$\big(\frac{g^2}{8\pi^2}\big)^{\!2}\frac{\zeta(3)}{2}$ and collecting all terms up to order $g^{2n+4}$, we finally obtain
\begin{align}
G_{\vec{n},\vec{m}}&=g^{2n}\,G^{(0)}_{\vec{n},\vec{m}}+g^{2n}\Big(\frac{g^2}{8\pi^2}\Big)^{\!2}3\,\zeta(3)
\Big[2n\big(C_\cR\,i_\cR-N^2\big)\,G^{(0)}_{\vec{n},\vec{m}}+\widehat{\cG}_{\vec{n},\vec{m}}\Big]\notag\\[2mm]
&=g^{2n} \,G_{\vec{n},\vec{m}}^{(0)} \,\bigg[ 1 +\Big(\frac{g^2}{8\pi^2}\Big)^{\!2}\, 3\zeta(3)\,\cC^{(2)}_{\vec{n},\vec{m}}+\cO(g^6)\bigg]
\label{result2MM}
\end{align}
where
\begin{equation}
		\cC^{(2)}_{\vec{n},\vec{m}}=2n\,\big(C_\cR\,i_\cR-N^2\big) +\frac{\widehat{\cG}_{\vec{n},\vec{m}}}{G_{\vec{n},\vec{m}}^{(0)}}~,
		\label{Cnm}
\end{equation}
which perfectly matches the result of the field theory computation presented in Section\,\ref{sec:Section 2} (see in particular Eq.s (\ref{Gnmren}) and (\ref{Gnmren1})). This highly non-trivial agreement shows that, within the regime (\ref{regime}),
the two-loop prediction of the matrix model generated by supersymmetric localization on $S^4$ yields the same results of the standard Feynman diagrams in $\mathbb{R}^4$ also for $\cN=2$ gauge theories with a non-zero $\beta$-function. A key point for this agreement is the identification of the running coupling constant at the separation scale in the two spaces, namely the distance $|x|$ in $\mathbb{R}^4$ between the two operators and the antipodal distance $2R$ in $S^4$, as indicated in Fig.\,\ref{FD}.
It is also interesting to observe that the numerical coefficient $3\zeta(3)$, which in the field-theory computations comes from the loop integrals in the limit $d\to4$, in the matrix model has instead a purely combinatorial explanation in terms of the symmetry factors of the effective diagrams (\ref{MM:diagramA},\ref{MM:diagramB}) produced by the quartic interaction $\frac{\zeta(3)}{2}\,\Tr^\prime a^{4}$.

Finally, we would like to stress that beyond the perturbative regime (\ref{regime}) we do not expect an exact correspondence between the results on $S^4$  and those in $\mathbb{R}^4$. In fact, when $\Lambda R\sim 1$, the observables receive power-like contributions of the form $C_n \left(\Lambda R\right)^n$, and it
is reasonable to expect that the coefficients $C_n$ of these \emph{infrared} corrections on the sphere and in flat space differ from each other. We also remark that supersymmetric localization can only provide information on a limited, though large, class of field-theory observables. For instance, it does not compute non-extremal correlators nor observables with complicated transcedental structures, as already happens in $\mathcal{N}=4$ SYM whose corresponding matrix model is purely Gaussian.

\section{Conclusions and perspectives}
\label{sec:conclusions}
In the perspective of extending the class of quantum field theories where exact results can be obtained, it is natural to pose the question of  whether supersymmetric localization techniques for protected observables in four-dimensional $\cN=2$ SYM theories on $S^4$ yield correct predictions for the same observables in flat space even when the $\beta$-function is non-zero.

Building on insights already present in the original work of Pestun \cite{Pestun:2007rz}, as well as on the  observations of \cite{Billo:2019job}, it was argued in \cite{Billo:2023igr, Billo:2024hvf, Billo:2024fst} that this is indeed the case. In these works it was shown that the expectation value of the $\frac{1}{2}$-BPS Wilson loop, computed up to order $g^6$, matches precisely the matrix model predictions arising from localization. In this perturbative correspondence, the coupling constant entering the matrix model is identified with the field-theoretic running coupling evaluated at the characteristic energy scale of the observable.

In this work, we extended this analysis to another class of protected observables: the two-point functions of chiral and anti-chiral operators. We explicitly verified that the matrix model reproduces exactly the field theory calculations up to the second non-trivial perturbative order. As in the case of the Wilson loop, we worked  in dimensional regularization, setting $d = 4 - 2\epsilon$, and carefully took into account the fact that that, in the $\cN=4$ theory, all loop corrections vanish in four dimensions and thus correspond at most to evanescent terms in the $\epsilon \to 0$ limit. Furthermore, we introduced a streamlined diagrammatic approach to the matrix model computation that closely resembles the structure of the field-theoretical Feynman expansion.

Based on the arguments and consistency checks presented above, we can conjecture
that the localization matrix model correctly captures protected observables within the energy regime specified by (\ref{regime}) even in $\cN=2$ SYM theories with massless matter and a non-zero $\beta$-function. This opens the door to using the matrix model as a tool for deriving new non-trivial results in such theories. We mention
two of them below.

\paragraph{Higher order diagrammatic computations:}
Computations in the matrix model are way easier than in field theory and one can push them to higher perturbative orders without much effort. For instance, in the diagrammatic approach, we can keep into account the sextic vertex in the interaction action (\ref{Seintis2}), namely
\begin{equation}
	\label{sextic}
	  \mathord{
		\begin{tikzpicture}[radius=2.cm, baseline=-0.65ex,scale=0.5]
			\begin{feynman}
				\vertex (A) at (2,0);
				\vertex (a1) at (1.5,1.5);
				\vertex (a2) at (-1.5,-1.5);
				\vertex (b1) at (-1.5,1.5);
				\vertex (b2) at (1.5,-1.5);
				\vertex (c1) at (1.8,1.8) {$a$};
				\vertex (c1) at (2.6,0) {$b$};
				\vertex (c1) at (1.8,-1.8) {$c$};
				\vertex (c1) at (-1.8,-1.8) {$d$};
				\vertex (c1) at (-2.6,0) {$e$};
				\vertex (c1) at (-1.8,1.8) {$f$};
				\vertex (B) at (-2,0);
				\vertex (C) at (0,2);
				\vertex (D) at (0,-2);
				\vertex (O) at (0,0);
				\diagram*{
					(A) --[ plain,thick] (O),
					(B)--[plain,thick] (O),
					(a1)--[plain,thick] (a2),
					(b1)--[plain,thick] (b2),
				};
			\end{feynman}
			\node at (0,0)[rectangle,fill,inner sep=3.1pt]{};
	\end{tikzpicture} }
	= \left(\frac{g^2}{8\pi^2}\right)^3\,\frac{\zeta(5)}{3}\, \widehat{C}_s^{abcdef}~,
\end{equation}
where $\widehat{C}_s^{abcdef} = \Tr^\prime\left(T^{(a}T^b T^c T^d T^e T^{f)}\right)$ is a completely symmetric tensor of rank six that generalizes the one in (\ref{Chatsymm}). The diagrams of order $g^6$ that contribute to a chiral/anti-chiral two-point function $\big\langle \cO_{\vec n}(a)\,\cO_{\vec m}(a)\big\rangle$ are the following:
\begin{equation}
	\label{diagramsg6mat}
		\mathord{
	\begin{tikzpicture}[scale=0.7, baseline=-0.65ex]
		\begin{feynman}
			\vertex (a) at (-2,0) {$\bullet$};
			\vertex (c) at (2,0) {$\bullet $ };
			\vertex (A) at (-2,0) ;
			\vertex (C) at (2,0) ;
			\vertex (o) at (0,0) {$.$};
			\vertex (o1) at (0,0.3) {$.$};
			\vertex (o2) at (0,-0.3) {$.$};
			\vertex (f) at (0,0.95) ;
			\vertex (f2) at (0,2.2)  ;
			\vertex (V) at (3.1,0); 
			\vertex (V) at (-3,0); 
			\diagram*{
				(A) -- [plain, half left, looseness=0.85] (C),
				(A) --[plain, half right, looseness=0.85] (C),
				(A) --[plain, half right, looseness=0.55] (C),
				(C) --[plain, half left, looseness=0.55] (f),
				(C) --[plain] (f),
				(A) --[plain] (f),
				(A) --[plain, half right, looseness=0.55] (f),
			};
		\end{feynman}
		\node at (0,0.96)[rectangle,fill,inner sep=3.1pt]{};
	\end{tikzpicture} 
	}
\quad	+ \quad
	\mathord{
		\begin{tikzpicture}[scale=0.7, baseline=-0.65ex]
			\begin{feynman}
				\vertex (a) at (-2,0) {$\bullet$};
				\vertex (c) at (2,0) {$\bullet $ };
				\vertex (A) at (-2,0) ;
				\vertex (C) at (2,0) ;
				\vertex (o) at (0,0) {$.$};
				\vertex (o1) at (0,0.3) {$.$};
				\vertex (o2) at (0,-0.3) {$.$};
				\vertex (f) at (0,0.95) ;
				\vertex (f2) at (0,2.2)  ;
				\vertex (V) at (3.1,0); 
				\vertex (V) at (-3,0); 
				\diagram*{
					(A) -- [plain, half left, looseness=0.85] (C),
					(A) --[plain, half right, looseness=0.85] (C),
					(A) --[plain, half right, looseness=0.55] (C),
					(f2) -- [plain,half left] (f),
					(f2) -- [plain,half right] (f),
					(C) --[plain, half left, looseness=0.55] (f),
					(A) --[plain, half right, looseness=0.55] (f),
				};
			\end{feynman}
			\node at (0,0.96)[rectangle,fill,inner sep=3.1pt]{};
		\end{tikzpicture} 
	} 
	\quad+\quad
		\mathord{
		\begin{tikzpicture}[scale=0.7, baseline=-0.65ex]
			\begin{feynman}
				\vertex (a) at (-2,0) {$\bullet$};
				\vertex (c) at (2,0) {$\bullet $ };
				\vertex (A) at (-2,0) ;
				\vertex (C) at (2,0) ;
				\vertex (o) at (0,0) {$.$};
				\vertex (o1) at (0,0.3) {$.$};
				\vertex (o2) at (0,-0.3) {$.$};
				\vertex (f) at (0,0.95) ;
				\vertex (f2) at (0.9,2.2)  ;
				\vertex (f3) at (-0.9,2.2)  ;
				\diagram*{
					(A) -- [plain, half left, looseness=0.85] (C),
					(A) --[plain, half right, looseness=0.85] (C),
					(A) --[plain, half right, looseness=0.5] (C),
					(A) --[plain, half left, looseness=0.5] (C),
					(f) -- [plain,half right,looseness=0.45] (f2),
					(f) -- [plain,half left,looseness=0.45] (f2),
					(f) -- [plain,half right,looseness=0.45] (f3),
					(f) -- [plain,half left,looseness=0.45] (f3),
				};
			\end{feynman}
			\node at (0,0.96)[rectangle,fill,inner sep=3.1pt]{};
		\end{tikzpicture} 
	}
\end{equation}
where the circular dots represent the two operators $\cO_{\vec n}$ and $\cO_{\vec m}$.
Keeping track of the combinatorics of the Wick contractions in the three diagrams, we get
\begin{align}
	\label{resg6}
	\Big(\frac{g^2}{8\pi^2}\Big)^{\!3} \zeta(5)\, n! 
	\bigg[& \frac{20}{3}n(n-1)(n-2)\, R_{\vec n}^{a_1 a_2 a_3 c_1 \ldots c_{n-3}}\, R_{\vec m}^{b_1 b_2 b_3 c_1 \ldots c_{n-3}} \,\widehat{C}_s^{a_1 a_2 a_3 b_1 b_2 b_3}
	\nonumber\\
	&~+ 30 n(n-1)\, R_{\vec n}^{a_1 a_2 c_1 \ldots c_{n-2}} \,R_{\vec m}^{b_1 b_2 c_1 \ldots c_{n-2}}\, 
	\widehat{C}_s^{a_1 a_2 b_1 b_2 d d}
	\nonumber\\
	&~+ 30 n\, R_{\vec n}^{a_1 c_1 \ldots c_{n-1}}\, R_{\vec m}^{b_1 c_1 \ldots c_{n-1}}\, 
	\widehat{C}_s^{a_1 b_1  d d e e}
	\bigg]~.
\end{align}
This matrix model prediction could possibly be checked against direct Feynman diagram computations in the superfield formalism for which a computerized approach suitable to this kind of correlators has been proposed in \cite{Bason:2023jiq}. It is however rather difficult to envisage Feynman diagram computations at even higher orders, while there is no particular problem to extend to higher orders the matrix model approach. 

Specializing the tensors $R_{\vec n}$ and $R_{\vec m}$, and evaluating the tensor $\widehat{C}_s$ for a given matter representation $\cR$, we can easily derive  explicit results. For instance, if we consider the low dimensional operators $\cO_2$ and $\cO_3$ in SQCD with $N_f$ flavors (and $N_a = N_s = 0$), from (\ref{G22ren}), (\ref{G33ren}) and the above formula (\ref{resg6}), we get   
\begin{align}
	\label{res22}
		G_{2,2} &=g^{4} \,\frac{N^2-1}{2}\,\bigg\{ 1 +\Big(\frac{g^2}{8\pi^2}\Big)^{\!2}\,\, 	\frac{3\zeta(3)}{2N}\,\big[N_f (2 N^2-3) -10 N^3\big] \nonumber\\
		& - \Big(\frac{g^2}{8\pi^2}\Big)^{\!3}\,\, \frac{5\zeta(5)}{N^2}\,
		\big[N_f(N^4- 3 N^2 + 3) - 14 N^5 \big] + O(g^8)\biggr\}~
\end{align}  
and 
\begin{align}
	\label{res33}
		G_{3,3} &=g^{6}\, \frac{3(N^2-1)(N^2-4)}{8N}\,
		\bigg\{ 1 +\Big(\frac{g^2}{8\pi^2}\Big)^{\!2}\,\, \frac{9\zeta(3)}{2N}\,
		\big[N_f (N^2-3)-4 N^3\big] \nonumber\\
		& - \Big(\frac{g^2}{8\pi^2}\Big)^{\!3}\,\, \frac{5\zeta(5)}{4N^2}\,
		\big[N_f(6N^4-31N^2+ 53) - 56 N^5\big]+ O(g^8)\biggr\}~.
\end{align}
Similar explicit expressions for higher dimensional operators can be easily worked out.
	
\paragraph{Large $N$ and strong coupling analysis:}
The matrix-model description leads, for certain classes of conformal $\cN=2$ theories, to powerful results in the large-$N$ 't Hooft limit in which the coupling $\lambda = g^2 N$ is kept fixed while $N$ tends to infinity \cite{Rey:2010ry,Passerini:2011fe,Bourgine:2011ie,Russo:2012ay,Baggio:2016skg,Rodriguez-Gomez:2016cem,Rodriguez-Gomez:2016ijh,Zarembo:2016bbk,Zarembo:2020tpf,Fiol:2020bhf,Beccaria:2020hgy,Beccaria:2021hvt}. In these cases, the $\lambda$-perturbative expansion of correlators involving single-trace operators can be resummed and even continued to strong coupling, allowing us to compare the result with  dual holographic descriptions \cite{Billo:2021rdb,Billo:2022xas, Billo:2022gmq, Billo:2022fnb}. In the light of the results we have obtained, it is natural to pose the question of whether such strong-coupling results can be extended to some classes of theories with non-zero $\beta$-function. Let us note that in the 't Hooft limit the coupling $g$ must be sent to zero in order to keep $\lambda$ constant when $N\to\infty$. The scale $1/R$ must therefore be much bigger than the IR scale $\Lambda$ at which the running coupling $g$ would diverge (see (\ref{regime})), and so one can neglect the instanton contributions and consider just the perturbative interaction action as we did here. This is a very interesting direction since it could lead to new insights regarding the holographic description of theories with a non-vanishing $\beta$-function. Work is currently underway along these lines \cite{next}.

\vskip 1.5cm
\noindent {\large {\bf Acknowledgments}}
\vskip 0.2cm
We would like to thank Marialuisa Frau, Francesco Galvagno, Gregory Korchemsky, Igor Pesando  and Paolo Vallarino for many useful discussions. 

The work of M.B and A.L. is partially supported by the MUR PRIN contract 2020-KR4KN2 ``String Theory as a bridge between Gauge Theories and Quantum Gravity'' and by the INFN project ST\&FI ``String Theory \& Fundamental Interactions''. The work of L.G. and A.T. is partially supported by the INFN project GAST ``Gauge And String Theory''.

\appendix
\section{One-loop diagrams with internal vertices}
\label{sec:pertcalc}
In this appendix we provide some details on the calculation of the diagram  
\begin{equation}
	\label{Sigma1def}
	\mathord{
		\begin{tikzpicture}[scale=0.5, baseline=-0.65ex]	
			\begin{feynman}
				\vertex (A) at (0,1.2);
				\vertex (B) at (0,-1.2);
				\vertex (a) at (-4.5,2) {$x, a_1$};
				\vertex (a1) at (-4,1.2) ;
				\vertex (b1) at (-1.1,1.2) ;
				\vertex (d) at (-4.5,-2.) {$x, a_2$};
				\vertex (d1) at (-4,-1.2) ;
				\vertex (c1) at (-1.1,-1.2) ;
				\vertex (e) at (4.5,2) {$0, b_1$};
				\vertex (e1) at (4,1.2) ;
				\vertex (f1) at (1.1,1.2) ;
				\vertex (h) at (4.5,-2.) {$0, b_2$};
				\vertex (h1) at (4,-1.2) ;
				\vertex (g1) at (1.1,-1.2) ;
				\diagram*{
					(b1) -- [plain] (e1),
					(c1) -- [plain] (h1),
					(a1)--[fermion] (b1),
					(d1)--[fermion] (c1),
					(f1)--[fermion] (e1),
					(g1)--[fermion] (h1),
					(A)--[photon] (B),
				};
			\end{feynman}
		\end{tikzpicture} 
	}  \equiv\, \Sigma^{a_1a_2b_1b_2}
\end{equation} 
which enters the one-loop correction (\ref{eq:1-loop}) of the two-point function. 
As already mentioned in the main text, we follow the notations and conventions of \cite{Billo:2017glv,Billo:2019job} to which we refer for further details. In particular we use the $\cN=1$ super-space formalism with the action and super-Feynman rules given in Section\,3.1 of \cite{Billo:2017glv}. Applying these rules, we find
\begin{align}
	\Sigma^{a_1a_2b_1b_2} &= -N g^2_B \,C^{a_1 a_2 b_1 b_2}  \!\!\int\! \dd z_3\dd z_4 \,\theta_{34}^2
	\, \bar{\theta}_{34}^2\, \mathbf{\Delta}(z_3,z_4) \mathbf{\Delta}(x,z_3) \mathbf{\Delta}(x,z_4) \mathbf{\Delta}(z_3,0)\mathbf{\Delta}(z_4,0)\notag\\[2mm]
	&\equiv g_B^2\, C^{a_1 a_2 b_1 b_2}\, \Sigma~.
	\label{eq:Sigma1 2ndstep}
\end{align}
Here $C$ is the tensor defined in (\ref{eq:C4A}) which encodes the color structure of the diagram, $z_3$ and
$z_4$ are the super-space points where the interaction vertices are inserted between which the vector multiplet is exchanged, and $\mathbf{\Delta}$ is the super-propagator. More explicitly, if we write the super-space points
as $z_i=(x_i,\theta_i,\bar{\theta}_i)$, the integration measure is $\dd z_i = \dd^4 x_i\, \dd^2\theta_i\,\dd^2\bar{\theta}_i$ and the super-propagator is
\begin{align}
\mathbf{\Delta}(z_i,z_j) = \mathrm{e}^{\ii(\theta_{i}\sigma\,\bar{\theta}_i+\theta_{j}\sigma\,\bar{\theta}_j-2\,
\theta_{i}\sigma\,\bar{\theta}_j)\cdot \partial_{x_{ij}}} \Delta(x_{ij},\epsilon)
\end{align}
where $\Delta$ is the scalar propagator (\ref{massless prop}). Furthermore, for the two external points where the scalar operators are inserted, $x$ and $0$, the corresponding fermionic coordinates are zero.
 
To proceed with the calculation of the function $\Sigma$ in (\ref{eq:Sigma1 2ndstep}), it is convenient to write the
propagators in momentum space using (\ref{massless prop}). In this way we find
\begin{equation}
	\label{eq:int rep Sigma1}
	\begin{split}
	\Sigma&=
-N \!\int\!\dd^2\theta_4\,\dd^2\bar{\theta}_4^2
\int\dfrac{\dd^d p}{(2\pi)^d}\dfrac{\dd^d k}{(2\pi)^d}\dfrac{\dd^d \ell}{(2\pi)^d}\,\dfrac{\ee^{-2\theta_4\,p\,\bar{\theta}_4}\,\ee^{\ii p x}}{ \ell^2 \, k^2 \,  (\ell-k)^2 \, (k+p)^2 \, (\ell+p)^2}\\[0.5em]
&=N \!\int\!\dfrac{\dd^d p}{(2\pi)^d}\dfrac{\dd^d k}{(2\pi)^d}\dfrac{\dd^d \ell}{(2\pi)^d}\,\dfrac{ p^2\, 
\ee^{\ii p x} }{ \ell^2 \,  k^2 \,  (\ell-k)^2 \, (k+p)^2 \, (\ell+p)^2}
\end{split}
\end{equation}
where to obtain the second line we have integrated over the Grassmann variables $\theta_4$ and $\bar{\theta}_4$ by Taylor expanding the exponential factor. The integrations over $k$ and $\ell$ can be performed using standard techniques \cite{Grozin:2005yg,Billo:2019job} and the result is proportional to $\zeta(3)$. Going through the calculation, we find
\begin{align}
\Sigma= e_{2,1}(x,\epsilon)\,\Delta^2(x,\epsilon)\quad
\mbox{with}\quad
e_{2,1}(x,\epsilon) = \frac{3 N \,\zeta (3) \,\epsilon \,(x^2)^{\epsilon } \,\Gamma (2-3 \epsilon )}{4^{1+2 \epsilon} \,\pi ^{2+\epsilon} \,\Gamma (1-\epsilon )^2\, \Gamma (1+2\epsilon)}
~.
\end{align}
Combining everything together, we see that (\ref{eq:Sigma1 2ndstep}) reproduces (\ref{v42Aform}).

\section{A few explicit formulas for the two-point correlators}
\label{App:Gmn}
Here we report the explicit expression of the renormalized correlators $G^*_{\vec{n},\vec{m}}$ defined in (\ref{Gnmren}) for operators of dimension $n=4$ and $5$ up to order $g^6$, when the matter content corresponds to the representation $\cR$ given in (\ref{eq:Ris}).

For the dimension 4 operators, we have
\begin{align}
G^*_{4,4}&=g^{8}\,\frac{(N^2\!-\!1)(N^4\!-\!6 N^2\!+\!18)}{4 N^2}\,\bigg\{ 1\!+\!
\Big(\frac{g^2}{8\pi^2}\Big)^{\!2}\,\frac{6\,\zeta(3)}{N(N^4\!-\!6 N^2\!+\!18)}\times\notag\\
&~~~~\times\Big[N_f\,(N^6\!-\!8 N^4\!+\!45 N^2\!-\!81)-4 N^7\!+\!12 N^5\!-\!54 N^3\label{G44}\\
&\qquad +N_a\,(N-2)(2 N^6\!-\!4 N^5\!-\!14 N^4\!+\!36 N^3\!+\!99 N^2\!-\!162 N\!-\!324)\notag\\
&\qquad +N_s\,(N+2)(2 N^6\!+\!4 N^5\!-\!14 N^4\!-\!36 N^3\!+\!99 N^2\!+\!162 N\!-\!324)\Big]\!+\!\cO(g^6)\bigg\}~,\notag
\end{align}
\begin{align}
G^*_{4,(2,2)}&=g^{8}\,\frac{(N^2\!-\!1)(2 N^2\!-\!3)}{2 N}\,\bigg\{ 1\!+\!\Big(\frac{g^2}{8\pi^2}\Big)^{\!2}\,
\frac{3\,\zeta(3)}{N(2N^2\!-\!3)}\times\notag\\
&~~~~\times\Big[N_f\,(5 N^4\!-\!18 N^2\!+\!27)\!+\!18 N^3\!-\!22 N^5 \label{G422}\\
&\qquad +N_a\,(N-2)(11 N^4\!-\!18 N^3\!-\!45 N^2\!+\!54 N\!+\!108)\notag\\
&\qquad +N_s\,(N+2)(11 N^4\!+\!18 N^3\!-\!45 N^2\!-\!54 N\!+\!108)\Big]+\cO(g^6)\bigg\}~,\notag
\end{align}
\begin{align}
G^*_{(2,2),(2,2)}&=g^{8}\,\frac{(N^4\!-\!1)}{2}\,\bigg\{ 1\!+\!\Big(\frac{g^2}{8\pi^2}\Big)^{\!2}\,\,
\frac{3\,\zeta(3)(N^2\!+\!3)}{N(N^2\!+\!1)}\times\notag\\
&~~~~\times\Big[N_f\,(2 N^2\!-\!3)-10 N^3+N_a\,(N-2)(5 N^2\!-\!6 N\!-\!12) \label{G2222}\\
&\qquad+N_s\,(N+2)(5 N^2\!+\!6 N\!-\!12)\Big]+\cO(g^6)\bigg\}~.\notag
\end{align}
For the dimension 5 operators, we have
\begin{align}
G^*_{5,5}&=g^{10}\,\frac{5(N^2\!-\!1) (N^2\!-\!4)(N^4\!+\!24)}{32 N^3}\,\bigg\{ 1\!+\!
\Big(\frac{g^2}{8\pi^2}\Big)^{\!2}\,\frac{15\,\zeta(3)}{2 N (N^4\!+\!24)}\times\notag\\
&~~~~\times\Big[N_f\,(N^6\!+\!N^4\!+\!24 N^2\!-\!144)-4 N^7\!-\!24 N^5\!-\!48 N^3\label{G55}\\
&\qquad +N_a\,(2 N^7\!-\!8 N^6\!+\!12 N^5\!-\!8 N^4\!+\!24 N^3\!-\!192 N^2\!+\!1152)
\notag\\
&\qquad +N_s\,(2 N^7\!+\!8 N^6\!+\!12 N^5\!+\!8 N^4\!+\!24 N^3\!+\!192 N^2\!-\!1152)\Big]+\cO(g^6)\bigg\}~,\notag
\end{align}
\begin{align}
G^*_{5,(3,2)}&=g^{10}\,\frac{15 (N^2\!-\!1) (N^2\!-\!2)(N^2\!-\!2)}{16 N^2}\,\bigg\{ 1\!+\!
\Big(\frac{g^2}{8\pi^2}\Big)^{\!2}\,\frac{3\,\zeta(3)}{N(N^2\!-\!2)}\times\notag\\
&~~~~\times\Big[N_f\,(N^4\!-\!14 N^2\!+\!30)+10 N^3 \!-\! 12 N^5\label{G532}\\
&\qquad +N_a\,(6 N^5\!-\!24 N^4\!-\!5 N^3\!+\!112 N^2\!-\!240)\notag\\
&\qquad +N_s\,(6 N^5\!+\!24 N^4\!-\!5 N^3\!-\!112 N^2\!+\!240)\Big]+\cO(g^6)\bigg\}~,\notag
\end{align}
\begin{align}
G^*_{(3,2),(3,2)}&=g^{10}\,\frac{3 (N^2\!-\!1)(N^2-\!4)(N^2\!+\!5)}{16 N}\,\bigg\{ 1\!+\!
\Big(\frac{g^2}{8\pi^2}\Big)^{\!2}\,\frac{3\,\zeta(3)}{2 N(N^2\!+\!5)}\times\notag\\
&~~~~\times\Big[N_f\,(5 N^4\!+\!37 N^2\!-\!150)-22 N^5\!-\!194 N^3\label{G3232}\\
&\qquad +N_a\,(11 N^5\!-\!40 N^4\!+\!97 N^3\!-\!296 N^2\!+\!1200)\notag\\
&\qquad +N_s\,(11 N^5\!+\!40 N^4\!+\!97 N^3\!+\!296 N^2\!-\!1200)\Big]+\cO(g^6)\bigg\}~.\notag
\end{align}
Explicit expressions for operators of higher dimensions can be easily worked out with a moderate computational effort.

\section{Identities for color factors}
\label{App:MM}
In this appendix we prove the relations (\ref{relationsMM}) obeyed by the color factors of the two-point correlators.

Let us first consider the amplitude (\ref{MM:diagramA}) which (up to an overall factor of $6\,g^{2n}$) is
\begin{align}
\cA_{\vec{n},\vec{m}}=
2n\,n!\,\widehat{C}_s^{a_1b_1 d \,d}\,\,R_{\vec n}^{a_1c_1\ldots c_{n-1}}\, R_{\vec m}^{b_1c_1 \ldots c_{n-1}}~.
\label{Anm}
\end{align}
Taking into account the definition (\ref{Chatsymm}) of the symmetric tensor $\widehat{C}_s$ and that of the
tensor $\widehat{C}$ given in (\ref{C4prime}) where we recognize a primed trace of the product four generators, it is straightforward to see that
\begin{align}
\cA_{\vec{n},\vec{m}}=\frac{2n\,n!}{3}\Big[\Tr^\prime T^{a_1}T^{b_1}T^{d}T^{d}\!+\Tr^\prime T^{a_1}T^{d}T^{b_1}T^{d}\!+\Tr^\prime T^{b_1}T^{a_1}T^{d}T^{d}\Big]R_{\vec n}^{a_1c_1\ldots c_{n-1}}R_{\vec m}^{b_1c_1 \ldots c_{n-1}}\,.
\label{Anm1}
\end{align}
In any representation $\cR$ we have the following identity
\begin{align}
\Tr_{\cR} T^{a}T^{b}T^{d}T^{d}=C_\cR \Tr_{\cR} T^{a}T^{b} =C_\cR\,i_\cR\,\delta^{ab}
\end{align}
where $C_\cR$ and $i_\cR$ are, respectively, the quadratic Casimir invariant and the Dynkin index of the representation $\cR$. Applying this identity to the primed trace, we have
\begin{align}
\Tr^\prime T^{a}T^{b}T^{d}T^{d}=\big(C_\cR\,i_\cR-N^2)\,\delta^{ab}
\label{identity1}
\end{align}
where we have used that for the adjoint representation $C_{\mathrm{adj}}=i_{\mathrm{adj}}=N$.

The second trace in the square brackets of (\ref{Anm1}) can be rewritten
as
\begin{align}
\Tr^\prime T^{a_1}T^{d}T^{b_1}T^{d}=\Tr^\prime T^{a_1}T^{b_1}T^{d}T^{d}+\ii\,f^{d\,b_1c}\,\Tr^\prime T^{a_1}T^{c}T^{d}
\label{identity2}
\end{align}
where $f^{d\,b_1c}$ are the structure constants. In any representation $\cR$ the following relation holds
\begin{align}
\Tr_{\cR} T^{a}T^{b}T^{c}=\ii\,\frac{i_\cR}{2}\,f^{abc}+x_\cR\,d^{abc}
\end{align}
where $x_\cR$ is a representation-dependent coefficient whose explicit expression is not needed and $d^{abc}$
are the symmetric $d$-symbols.
From this relation, it follows that
\begin{align}
\Tr^\prime T^{a}T^{b}T^{c}=\ii\,\frac{(i_\cR-N)}{2}\,f^{abc}+x^\prime\,d^{abc}=-\ii\,
\frac{\beta_0}{4}\,f^{abc}+x^\prime\,d^{abc}
\label{trace3}
\end{align}
where in the last step we have used that $\beta_0=2(N-i_\cR)$. When this result is inserted in (\ref{identity2}), only the part proportional to the structure constants survives and so one finds
\begin{align}
\Tr^\prime T^{a_1}T^{d}T^{b_1}T^{d}=\Tr^\prime T^{a_1}T^{b_1}T^{d}T^{d}+\frac{\beta_0}{4}\,f^{d\,b_1c}f^{a_1cd} 
=\Big[C_\cR\,i_\cR-N^2+\frac{N\beta_0}{4}\Big]\,\delta^{a_1b_1}
\label{identity3}
\end{align}
where to write the last equality we have used (\ref{identity1}) and $f^{abc}f^{abd}=N\,\delta^{cd}$.

Using (\ref{identity1}) and (\ref{identity3}) in (\ref{Anm1}), it is straightforward to obtain
\begin{align}
\cA_{\vec{n},\vec{m}}&=\frac{2n\,n!}{3}\Big[3\big(C_\cR\,i_\cR-N^2\big)+\frac{N\beta_0}{4}\Big]\,\delta^{a_1b_1}\,R_{\vec n}^{a_1c_1\ldots c_{n-1}}R_{\vec m}^{b_1c_1 \ldots c_{n-1}}\notag\\[2mm]
&=2n\,\Big(C_\cR\,i_\cR-N^2+\frac{N \beta_0}{12}\Big)\,G^{(0)}_{\vec{n},\vec{m}}
\label{identity4}
\end{align}
which is precisely the relation given in (\ref{relationA}).

Let us now consider the amplitude (\ref{MM:diagramB}) which (again up to an overall factor of $6\,g^{2n}$) is
\begin{align}
\cB_{\vec{n},\vec{m}}=n(n-1)\,n!\,\widehat{C}_s^{a_1a_2b_1b_2}\,\,R_{\vec n}^{a_1a_2c_1\ldots c_{n-2}}\, R_{\vec m}^{b_1b_2c_1 \ldots c_{n-2}}~.
\label{Bnm}
\end{align}
Proceeding as in the previous case, from (\ref{Chatsymm}) we get
\begin{align}
\cB_{\vec{n},\vec{m}}=\frac{n(n-1)\,n!}{3}\Big[\Tr^\prime T^{a_1}T^{b_1}T^{a_2}T^{b_2}+2
\Tr^\prime T^{a_1}T^{a_2}T^{b_1}T^{b_2}\Big] R_{\vec n}^{a_1a_2c_1\ldots c_{n-2}}\, R_{\vec m}^{b_1b_2c_1 \ldots c_{n-2}}\,.
\end{align}
The first primed trace in the above square brackets is just the tensor $\widehat{C}^{a_1 a_2 b_1 b_2}$ defined in (\ref{C4prime}), while the second primed trace can be manipulated as follows
\begin{align}
\Tr^\prime T^{a_1}T^{a_2}T^{b_1}T^{b_2}&=\Tr^\prime T^{a_1}T^{b_1}T^{a_2}T^{b_2}+\ii f^{a_2b_1c}\,\Tr^\prime
 T^{a_1}T^{c}T^{b_2} \notag\\
 &=\widehat{C}^{a_1 a_2 b_1 b_2}-\frac{\beta_0}{4}\,f^{a_2b_1c}f^{a_1c\,b_2}+\ii\,x^\prime f^{a_2b_1c}d^{a_1c\,b_2}\notag\\
 &=\widehat{C}^{a_1 a_2 b_1 b_2}+\frac{N\beta_0}{4}\,C^{a_1 a_2 b_1 b_2}+\ii\,x^\prime f^{a_2b_1c}d^{a_1c\,b_2}
 \label{identity5}
\end{align}
where the second line is a consequence of (\ref{trace3}) and in the third line we have introduced the tensor $C$ defined in (\ref{eq:C4A}). Notice that the part proportional to $x^\prime$ can be discarded since it gives rise to structures that are anti-symmetric in $(a_1,a_2)$ and $(b_1,b_2)$ and thus vanish when contracted with the symmetric tensors $R_{\vec{n}}$ and $R_{\vec{m}}$.

Using the identity (\ref{identity5}) in (\ref{Bnm}), we obtain
\begin{align}
\cB_{\vec{n},\vec{m}}&=\frac{n(n-1)\,n!}{3}\Big[3\,\widehat{C}^{a_1 a_2 b_1 b_2}+\frac{N\beta_0}{2}\,C^{a_1 a_2 b_1 b_2}\Big]R_{\vec n}^{a_1a_2c_1\ldots c_{n-2}}\, R_{\vec m}^{b_1b_2c_1 \ldots c_{n-2}}\notag\\
&=n(n-1)\,n!\,\widehat{C}^{a_1 a_2 b_1 b_2}\,\,R_{\vec n}^{a_1a_2c_1\ldots c_{n-2}}\, R_{\vec m}^{b_1b_2c_1 \ldots c_{n-2}}\notag\\
&\qquad+\frac{N\beta_0}{6}\,n(n-1)\,n!\,C^{a_1 a_2 b_1 b_2}\,\,R_{\vec n}^{a_1a_2c_1\ldots c_{n-2}}\, R_{\vec m}^{b_1b_2c_1 \ldots c_{n-2}}~.
\end{align}
The term proportional to $\widehat{C}$ is exactly $\widehat{\cG}_{\vec{n},\vec{m}}$ defined in (\ref{Gprime}), while the term proportional to $C$ corresponds to the color factor of the diagram (\ref{v42Acontribution}). The latter can be evaluated using the fusion/fission identities as described in \cite{Billo:2019job} which give
\begin{align}
n(n-1)\,n!\,C^{a_1 a_2 b_1 b_2}\,\,R_{\vec n}^{a_1a_2c_1\ldots c_{n-2}}\, R_{\vec m}^{b_1b_2c_1 \ldots c_{n-2}}=-n\,G^{(0)}_{\vec{n},\vec{m}}~.
\end{align}
Therefore, collecting all terms we have 
\begin{align}
\cB_{\vec{n},\vec{m}}=\widehat{\cG}_{\vec{n},\vec{m}}-n\,\frac{N\beta_0}{6}\,G^{(0)}_{\vec{n},\vec{m}}
\end{align}
which is the relation written in (\ref{relationB}).


\begin{thebibliography}{10}

\bibitem{Pestun:2016zxk}
V.~Pestun et~al., {\it {Localization techniques in quantum field theories}},
  {\em J. Phys. A} {\bf 50} (2017), no.~44 440301,
  [\href{http://arxiv.org/abs/1608.02952}{{\tt arXiv:1608.02952}}].

\bibitem{Pestun:2007rz}
V.~Pestun, {\it {Localization of gauge theory on a four-sphere and
  supersymmetric Wilson loops}},  {\em Commun. Math. Phys.} {\bf 313} (2012)
  71--129, [\href{http://arxiv.org/abs/0712.2824}{{\tt arXiv:0712.2824}}].

\bibitem{Baggio:2014ioa}
M.~Baggio, V.~Niarchos, and K.~Papadodimas, {\it {tt$^{*}$ equations,
  localization and exact chiral rings in 4d $ \mathcal{N} $ =2 SCFTs}},  {\em
  JHEP} {\bf 02} (2015) 122, [\href{http://arxiv.org/abs/1409.4212}{{\tt
  arXiv:1409.4212}}].

\bibitem{Baggio:2014sna}
M.~Baggio, V.~Niarchos, and K.~Papadodimas, {\it {Exact correlation functions
  in $SU(2) \mathcal N=2$ superconformal QCD}},  {\em Phys. Rev. Lett.} {\bf
  113} (2014), no.~25 251601, [\href{http://arxiv.org/abs/1409.4217}{{\tt
  arXiv:1409.4217}}].

\bibitem{Gerchkovitz:2016gxx}
E.~Gerchkovitz, J.~Gomis, N.~Ishtiaque, A.~Karasik, Z.~Komargodski, and S.~S.
  Pufu, {\it {Correlation Functions of Coulomb Branch Operators}},  {\em JHEP}
  {\bf 01} (2017) 103, [\href{http://arxiv.org/abs/1602.05971}{{\tt
  arXiv:1602.05971}}].

\bibitem{Rodriguez-Gomez:2016cem}
D.~Rodriguez-Gomez and J.~G. Russo, {\it {Operator mixing in large $N$
  superconformal field theories on S$^{4}$ and correlators with Wilson loops}},
   {\em JHEP} {\bf 12} (2016) 120, [\href{http://arxiv.org/abs/1607.07878}{{\tt
  arXiv:1607.07878}}].

\bibitem{Rodriguez-Gomez:2016ijh}
D.~Rodriguez-Gomez and J.~G. Russo, {\it {Large N Correlation Functions in
  Superconformal Field Theories}},  {\em JHEP} {\bf 06} (2016) 109,
  [\href{http://arxiv.org/abs/1604.07416}{{\tt arXiv:1604.07416}}].

\bibitem{Billo:2017glv}
M.~Bill\`o, F.~Fucito, A.~Lerda, J.~F. Morales, Y.~S. Stanev, and C.~Wen, {\it
  {Two-point correlators in $N =2$ gauge theories}},  {\em Nucl. Phys. B} {\bf
  926} (2018) 427--466, [\href{http://arxiv.org/abs/1705.02909}{{\tt
  arXiv:1705.02909}}].

\bibitem{Billo:2018oog}
M.~Bill\`o, F.~Galvagno, P.~Gregori, and A.~Lerda, {\it {Correlators between
  Wilson loop and chiral operators in $ \mathcal{N}=2 $ conformal gauge
  theories}},  {\em JHEP} {\bf 03} (2018) 193,
  [\href{http://arxiv.org/abs/1802.09813}{{\tt arXiv:1802.09813}}].

\bibitem{Galvagno:2020cgq}
F.~Galvagno and M.~Preti, {\it {Chiral correlators in $ \mathcal{N} $ = 2
  superconformal quivers}},  {\em JHEP} {\bf 05} (2021) 201,
  [\href{http://arxiv.org/abs/2012.15792}{{\tt arXiv:2012.15792}}].

\bibitem{Beccaria:2020hgy}
M.~Beccaria, M.~Bill\`o, F.~Galvagno, A.~Hasan, and A.~Lerda, {\it {$
  \mathcal{N} $ = 2 Conformal SYM theories at large $ \mathcal{N} $}},  {\em
  JHEP} {\bf 09} (2020) 116, [\href{http://arxiv.org/abs/2007.02840}{{\tt
  arXiv:2007.02840}}].

\bibitem{Fiol:2021icm}
B.~Fiol and A.~R. Fukelman, {\it {The planar limit of $ \mathcal{N} $ = 2
  chiral correlators}},  {\em JHEP} {\bf 08} (2021) 032,
  [\href{http://arxiv.org/abs/2106.04553}{{\tt arXiv:2106.04553}}].

\bibitem{Drukker:2007qr}
N.~Drukker, S.~Giombi, R.~Ricci, and D.~Trancanelli, {\it {Supersymmetric
  Wilson loops on S**3}},  {\em JHEP} {\bf 05} (2008) 017,
  [\href{http://arxiv.org/abs/0711.3226}{{\tt arXiv:0711.3226}}].

\bibitem{Bassetto:2008yf}
A.~Bassetto, L.~Griguolo, F.~Pucci, and D.~Seminara, {\it {Supersymmetric
  Wilson loops at two loops}},  {\em JHEP} {\bf 06} (2008) 083,
  [\href{http://arxiv.org/abs/0804.3973}{{\tt arXiv:0804.3973}}].

\bibitem{Pestun:2009nn}
V.~Pestun, {\it {Localization of the four-dimensional N=4 SYM to a two-sphere
  and 1/8 BPS Wilson loops}},  {\em JHEP} {\bf 12} (2012) 067,
  [\href{http://arxiv.org/abs/0906.0638}{{\tt arXiv:0906.0638}}].

\bibitem{Giombi:2009ds}
S.~Giombi and V.~Pestun, {\it {Correlators of local operators and 1/8 BPS
  Wilson loops on S**2 from 2d YM and matrix models}},  {\em JHEP} {\bf 10}
  (2010) 033, [\href{http://arxiv.org/abs/0906.1572}{{\tt arXiv:0906.1572}}].

\bibitem{Andree:2010na}
R.~Andree and D.~Young, {\it {Wilson Loops in N=2 Superconformal Yang-Mills
  Theory}},  {\em JHEP} {\bf 09} (2010) 095,
  [\href{http://arxiv.org/abs/1007.4923}{{\tt arXiv:1007.4923}}].

\bibitem{Billo:2019fbi}
M.~Bill\`o, F.~Galvagno, and A.~Lerda, {\it {BPS wilson loops in generic
  conformal $ \mathcal{N} $ = 2 SU(N) SYM theories}},  {\em JHEP} {\bf 08}
  (2019) 108, [\href{http://arxiv.org/abs/1906.07085}{{\tt arXiv:1906.07085}}].

\bibitem{Beccaria:2021vuc}
M.~Beccaria, G.~V. Dunne, and A.~A. Tseytlin, {\it {BPS Wilson loop in $
  \mathcal{N} $ = 2 superconformal SU(N)
  \textquotedblleft{}orientifold\textquotedblright{} gauge theory and
  weak-strong coupling interpolation}},  {\em JHEP} {\bf 07} (2021) 085,
  [\href{http://arxiv.org/abs/2104.12625}{{\tt arXiv:2104.12625}}].

\bibitem{Komatsu:2020sup}
S.~Komatsu and Y.~Wang, {\it {Non-perturbative defect one-point functions in
  planar $\mathcal{N}=4$ super-Yang-Mills}},  {\em Nucl. Phys. B} {\bf 958}
  (2020) 115120, [\href{http://arxiv.org/abs/2004.09514}{{\tt
  arXiv:2004.09514}}].

\bibitem{Beccaria:2022bjo}
M.~Beccaria and A.~Cabo-Bizet, {\it {1/N expansion of the D3-D5 defect CFT at
  strong coupling}},  {\em JHEP} {\bf 02} (2023) 208,
  [\href{http://arxiv.org/abs/2212.12415}{{\tt arXiv:2212.12415}}].

\bibitem{Correa:2012at}
D.~Correa, J.~Henn, J.~Maldacena, and A.~Sever, {\it {An exact formula for the
  radiation of a moving quark in N=4 super Yang Mills}},  {\em JHEP} {\bf 06}
  (2012) 048, [\href{http://arxiv.org/abs/1202.4455}{{\tt arXiv:1202.4455}}].

\bibitem{Bonini:2015fng}
M.~Bonini, L.~Griguolo, M.~Preti, and D.~Seminara, {\it {Bremsstrahlung
  function, leading L\"uscher correction at weak coupling and localization}},
  {\em JHEP} {\bf 02} (2016) 172, [\href{http://arxiv.org/abs/1511.05016}{{\tt
  arXiv:1511.05016}}].

\bibitem{Fiol:2015spa}
B.~Fiol, E.~Gerchkovitz, and Z.~Komargodski, {\it {Exact Bremsstrahlung
  Function in $N=2$ Superconformal Field Theories}},  {\em Phys. Rev. Lett.}
  {\bf 116} (2016), no.~8 081601, [\href{http://arxiv.org/abs/1510.01332}{{\tt
  arXiv:1510.01332}}].

\bibitem{Mitev:2015oty}
V.~Mitev and E.~Pomoni, {\it {Exact Bremsstrahlung and Effective Couplings}},
  {\em JHEP} {\bf 06} (2016) 078, [\href{http://arxiv.org/abs/1511.02217}{{\tt
  arXiv:1511.02217}}].

\bibitem{Bianchi:2019dlw}
L.~Bianchi, M.~Bill\`o, F.~Galvagno, and A.~Lerda, {\it {Emitted Radiation and
  Geometry}},  {\em JHEP} {\bf 01} (2020) 075,
  [\href{http://arxiv.org/abs/1910.06332}{{\tt arXiv:1910.06332}}].
  
  \bibitem{Binder:2019jwn}
D.~J.~Binder, S.~M.~Chester, S.~S.~Pufu and Y.~Wang,
{\it {$ \mathcal{N} $ = 4 Super-Yang-Mills correlators at strong coupling from string theory and localization}},  {\em JHEP} {\bf 12} (2019) 119, [\href{http://arxiv.org/abs/1902.06263}{{\tt arXiv:1902.06263}}].

\bibitem{Chester:2019jas}
S.~M.~Chester, M.~B.~Green, S.~S.~Pufu, Y.~Wang and C.~Wen,
{\it {Modular invariance in superstring theory from $ \mathcal{N} $ = 4 super-Yang-Mills}},  {\em JHEP} {\bf 11} (2020) 016, [\href{http://arxiv.org/abs/1912.13365}{{\tt arXiv:1912.13365}}].

\bibitem{Chester:2020dja}
S.~M.~Chester and S.~S.~Pufu,
{\it {Far beyond the planar limit in strongly-coupled $ \mathcal{N} $ = 4 SYM}},  {\em JHEP} {\bf 01} (2021) 103, [\href{http://arxiv.org/abs/2003.08412}{{\tt arXiv:2003.08412}}].

\bibitem{Dorigoni:2021guq}
D.~Dorigoni, M.~B.~Green and C.~Wen,
{\it {Exact properties of an integrated correlator in $ \mathcal{N} $ = 4 SU(N) SYM}},  {\em JHEP} {\bf 05} (2021) 089, [\href{http://arxiv.org/abs/2102.09537}{{\tt arXiv:2102.09537}}].

\bibitem{Paul:2022piq}
H.~Paul, E.~Perlmutter and H.~Raj,
{\it {Integrated correlators in $ \mathcal{N} $ = 4 SYM via SL(2, {\ensuremath{\mathbb{Z}}}) spectral theory}},  {\em JHEP} {\bf 01} (2023) 149, [\href{http://arxiv.org/abs/2209.06639}{{\tt arXiv:2102.09537}}].

\bibitem{Alday:2023pet}
L.~F.~Alday, S.~M.~Chester, D.~Dorigoni, M.~B.~Green and C.~Wen,
{\it {Relations between integrated correlators in $ \mathcal{N} $ = 4 supersymmetric Yang-Mills theory}},  {\em JHEP} {\bf 05} (2024) 044, [\href{http://arxiv.org/abs/2310.12322}{{\tt arXiv:2310.12322}}].

\bibitem{Brown:2024tru}
A.~Brown, F.~Galvagno and C.~Wen,
{\it {Exact results for giant graviton four-point correlators}},  {\em JHEP} {\bf 07} (2024) 049, [\href{http://arxiv.org/abs/2403.17263}{{\tt arXiv:2403.17263}}].

\bibitem{Behan:2023fqq}
C.~Behan, S.~M.~Chester and P.~Ferrero,
{\it {Gluon scattering in AdS at finite string coupling from localization}},  {\em JHEP} {\bf 02} (2024) 042, [\href{http://arxiv.org/abs/2305.01016}{{\tt arXiv:2305.01016}}].

\bibitem{Billo:2023kak}
M.~Billo, M.~Frau, A.~Lerda and A.~Pini,
{\it {A matrix-model approach to integrated correlators in a $ \mathcal{N} $ = 2 SYM theory}},  {\em JHEP} {\bf 01} (2024) 154, [\href{http://arxiv.org/abs/2311.17178}{{\tt arXiv:2311.17178}}].

\bibitem{Pini:2024uia}
A.~Pini and P.~Vallarino,
{\it {Integrated correlators at strong coupling in an orbifold of $ \mathcal{N} $ = 4 SYM}},  {\em JHEP} {\bf 06} (2024) 170, [\href{http://arxiv.org/abs/2404.03466}{{\tt arXiv:2404.03466}}].

\bibitem{Billo:2024ftq}
M.~Billo, M.~Frau, A.~Lerda, A.~Pini and P.~Vallarino,
{\it {Integrated correlators in a $ \mathcal{N} $ = 2 SYM theory with fundamental flavors: a matrix-model perspective}},  {\em JHEP} {\bf 11} (2024) 172, [\href{http://arxiv.org/abs/2407.03509}{{\tt arXiv:2407.03509}}].

\bibitem{Pini:2024zwi}
A.~Pini,
{\it {Integrated correlators with a Wilson line in a $ \mathcal{N} $ = 2 quiver gauge theory at strong coupling}},  {\em JHEP} {\bf 01} (2025) 195, [\href{http://arxiv.org/abs/2410.17342}{{\tt arXiv:2410.17342}}].

\bibitem{DeLillo:2025hal}
L.~De Lillo, M.~Frau and A.~Pini,
{\it {Integrated line-defect correlators in Sp(N) SCFTs at strong coupling}},  {\em JHEP} {\bf 06} (2025) 078, [\href{http://arxiv.org/abs/2503.04902}{{\tt arXiv:2503.04902}}].

\bibitem{Beccaria:2021ism}
M.~Beccaria, G.~V. Dunne, and A.~A. Tseytlin, {\it {Strong coupling expansion
  of free energy and BPS Wilson loop in $ \mathcal{N} $ = 2 superconformal
  models with fundamental hypermultiplets}},  {\em JHEP} {\bf 08} (2021) 102,
  [\href{http://arxiv.org/abs/2105.14729}{{\tt arXiv:2105.14729}}].

\bibitem{Beccaria:2021hvt}
M.~Beccaria, M.~Bill\`o, M.~Frau, A.~Lerda, and A.~Pini, {\it {Exact results in
  a $ \mathcal{N} $ = 2 superconformal gauge theory at strong coupling}},  {\em
  JHEP} {\bf 07} (2021) 185, [\href{http://arxiv.org/abs/2105.15113}{{\tt
  arXiv:2105.15113}}].

\bibitem{Billo:2021rdb}
M.~Bill\`o, M.~Frau, F.~Galvagno, A.~Lerda, and A.~Pini, {\it {Strong-coupling
  results for $ \mathcal{N} $ = 2 superconformal quivers and holography}},
  {\em JHEP} {\bf 10} (2021) 161, [\href{http://arxiv.org/abs/2109.00559}{{\tt
  arXiv:2109.00559}}].

\bibitem{Billo:2022gmq}
M.~Bill\`o, M.~Frau, A.~Lerda, A.~Pini, and P.~Vallarino, {\it {Structure
  Constants in N=2 Superconformal Quiver Theories at Strong Coupling and
  Holography}},  {\em Phys. Rev. Lett.} {\bf 129} (2022), no.~3 031602,
  [\href{http://arxiv.org/abs/2206.13582}{{\tt arXiv:2206.13582}}].

\bibitem{Billo:2022fnb}
M.~Bill\`o, M.~Frau, A.~Lerda, A.~Pini, and P.~Vallarino, {\it {Localization vs
  holography in 4d$ \mathcal{N} $ = 2 quiver theories}},  {\em JHEP} {\bf 10}
  (2022) 020, [\href{http://arxiv.org/abs/2207.08846}{{\tt arXiv:2207.08846}}].

\bibitem{Beccaria:2022ypy}
M.~Beccaria, G.~P. Korchemsky, and A.~A. Tseytlin, {\it {Strong coupling
  expansion in $ \mathcal{N} $ superconformal theories and the Bessel kernel}},
   {\em JHEP} {\bf 09} (2022) 226, [\href{http://arxiv.org/abs/2207.11475}{{\tt
  arXiv:2207.11475}}].

\bibitem{Beccaria:2023kbl}
M.~Beccaria, G.~P. Korchemsky, and A.~A. Tseytlin, {\it {Non-planar corrections
  in orbifold/orientifold $ \mathcal{N} $ = 2 superconformal theories from
  localization}},  {\em JHEP} {\bf 05} (2023) 165,
  [\href{http://arxiv.org/abs/2303.16305}{{\tt arXiv:2303.16305}}].

\bibitem{Billo:2022xas}
M.~Bill\`o, M.~Frau, A.~Lerda, A.~Pini, and P.~Vallarino, {\it {Three-point
  functions in a $ \mathcal{N} $ = 2 superconformal gauge theory and their
  strong-coupling limit}},  {\em JHEP} {\bf 08} (2022) 199,
  [\href{http://arxiv.org/abs/2202.06990}{{\tt arXiv:2202.06990}}].

\bibitem{Pini:2023svd}
A.~Pini and P.~Vallarino, {\it {Defect correlators in a $ \mathcal{N} $ = 2
  SCFT at strong coupling}},  {\em JHEP} {\bf 06} (2023) 050,
  [\href{http://arxiv.org/abs/2303.08210}{{\tt arXiv:2303.08210}}].

\bibitem{Pini:2023lyo}
A.~Pini and P.~Vallarino, {\it {Wilson loop correlators at strong coupling in $
  \mathcal{N} $ = 2 quiver gauge theories}},  {\em JHEP} {\bf 11} (2023) 003,
  [\href{http://arxiv.org/abs/2308.03848}{{\tt arXiv:2308.03848}}].

\bibitem{Korchemsky:2025eyc}
G.~P. Korchemsky and A.~Testa, {\it {Correlation functions in four-dimensional
  superconformal long circular quivers}},
  \href{http://arxiv.org/abs/2501.17223}{{\tt arXiv:2501.17223}}.

\bibitem{Belitsky:2020hzs}
A.~V. Belitsky and G.~P. Korchemsky, {\it {Circular Wilson loop in N=2* super
  Yang-Mills theory at two loops and localization}},  {\em JHEP} {\bf 04}
  (2021) 089, [\href{http://arxiv.org/abs/2003.10448}{{\tt arXiv:2003.10448}}].

\bibitem{Billo:2019job}
M.~Bill\`o, F.~Fucito, G.~P. Korchemsky, A.~Lerda, and J.~F. Morales, {\it
  {Two-point correlators in non-conformal $ \mathcal{N} $ = 2 gauge theories}},
   {\em JHEP} {\bf 05} (2019) 199, [\href{http://arxiv.org/abs/1901.09693}{{\tt
  arXiv:1901.09693}}].

\bibitem{Billo:2023igr}
M.~Bill\`o, L.~Griguolo, and A.~Testa, {\it {Remarks on BPS Wilson loops in
  non-conformal $ \mathcal{N} $ = 2 gauge theories and localization}},  {\em
  JHEP} {\bf 01} (2024) 160, [\href{http://arxiv.org/abs/2311.17692}{{\tt
  arXiv:2311.17692}}].

\bibitem{Billo:2024fst}
M.~Bill\`o, L.~Griguolo, and A.~Testa, {\it {1/2 BPS Wilson loops in
  non-conformal N = 2 gauge theories and localization: a three-loop analysis}},
   \href{http://arxiv.org/abs/2410.14847}{{\tt arXiv:2410.14847}}.

\bibitem{Billo:2024hvf}
M.~Bill\`o, L.~Griguolo, and A.~Testa, {\it {Supersymmetric localization and
  non-conformal $\mathcal{N}=2$ SYM theories in the perturbative regime}},
  \href{http://arxiv.org/abs/2407.11222}{{\tt arXiv:2407.11222}}.

\bibitem{Erickson:2000af}
J.~K. Erickson, G.~W. Semenoff, and K.~Zarembo, {\it {Wilson loops in N=4
  supersymmetric Yang-Mills theory}},  {\em Nucl. Phys. B} {\bf 582} (2000)
  155--175, [\href{http://arxiv.org/abs/hep-th/0003055}{{\tt hep-th/0003055}}].

\bibitem{Bianchi:2023llc}
M.~S.~Bianchi, {\it {Protected and uniformly transcendental}},  {\em
  JHEP} {\bf 09} (2023) 121, [\href{http://arxiv.org/abs/2306.06239}{{\tt
  arXiv:2306.06239}}].

\bibitem{Bason:2023jiq}
D.~Bason and M.~Bill\`o, {\it {$\theta $-diagram technique for ${\mathcal
  {N}}=1$, $d=4$ superfields}},  {\em Eur. Phys. J. C} {\bf 83} (2023), no.~10
  892, [\href{http://arxiv.org/abs/2301.11717}{{\tt arXiv:2301.11717}}].

\bibitem{Rey:2010ry}
S.-J. Rey and T.~Suyama, {\it {Exact Results and Holography of Wilson Loops in
  N=2 Superconformal (Quiver) Gauge Theories}},  {\em JHEP} {\bf 01} (2011)
  136, [\href{http://arxiv.org/abs/1001.0016}{{\tt arXiv:1001.0016}}].

\bibitem{Passerini:2011fe}
F.~Passerini and K.~Zarembo, {\it {Wilson Loops in N=2 Super-Yang-Mills from
  Matrix Model}},  {\em JHEP} {\bf 09} (2011) 102,
  [\href{http://arxiv.org/abs/1106.5763}{{\tt arXiv:1106.5763}}]. [Erratum:
  JHEP 10, 065 (2011)].

\bibitem{Bourgine:2011ie}
J.-E. Bourgine, {\it {A Note on the integral equation for the Wilson loop in N
  = 2 D=4 superconformal Yang-Mills theory}},  {\em J. Phys. A} {\bf 45} (2012)
  125403, [\href{http://arxiv.org/abs/1111.0384}{{\tt arXiv:1111.0384}}].

\bibitem{Russo:2012ay}
J.~G. Russo and K.~Zarembo, {\it {Large N Limit of N=2 SU(N) Gauge Theories
  from Localization}},  {\em JHEP} {\bf 10} (2012) 082,
  [\href{http://arxiv.org/abs/1207.3806}{{\tt arXiv:1207.3806}}].

\bibitem{Baggio:2016skg}
M.~Baggio, V.~Niarchos, K.~Papadodimas, and G.~Vos, {\it {Large-N correlation
  functions in $ \mathcal{N} $ = 2 superconformal QCD}},  {\em JHEP} {\bf 01}
  (2017) 101, [\href{http://arxiv.org/abs/1610.07612}{{\tt arXiv:1610.07612}}].

\bibitem{Zarembo:2016bbk}
K.~Zarembo, {\it {Localization and AdS/CFT Correspondence}},  {\em J. Phys. A}
  {\bf 50} (2017), no.~44 443011, [\href{http://arxiv.org/abs/1608.02963}{{\tt
  arXiv:1608.02963}}].

\bibitem{Zarembo:2020tpf}
K.~Zarembo, {\it {Quiver CFT at strong coupling}},  {\em JHEP} {\bf 06} (2020)
  055, [\href{http://arxiv.org/abs/2003.00993}{{\tt arXiv:2003.00993}}].

\bibitem{Fiol:2020bhf}
B.~Fiol, J.~Mart\'\i{}nez-Montoya, and A.~Rios~Fukelman, {\it {The planar limit
  of $\mathcal{N}=2$ superconformal field theories}},  {\em JHEP} {\bf 05}
  (2020) 136, [\href{http://arxiv.org/abs/2003.02879}{{\tt arXiv:2003.02879}}].


\bibitem{next}
M.~Bill\`o, L.~Griguolo, A.~Lerda, and A.~Testa, {\it {in preparation.}}

\bibitem{Grozin:2005yg}
A.~Grozin, {\it {Lectures on QED and QCD}},  in {\em {3rd Dubna International
  Advanced School of Theoretical Physics}}, 8, 2005.
\newblock \href{http://arxiv.org/abs/hep-ph/0508242}{{\tt hep-ph/0508242}}.

\end{thebibliography}

\providecommand{\href}[2]{#2}\begingroup\raggedright\endgroup

\end{document}